\documentclass[aps,prb,twocolumn,superscriptaddress,footinbib]{revtex4-1}
\usepackage{graphicx}
\usepackage{multirow}
\usepackage{color}
\usepackage{amsmath}
\usepackage{amsfonts}
\usepackage{amssymb}
\usepackage{graphicx}
\graphicspath{{Figures/}}
\usepackage{epstopdf}
\usepackage{empheq}
\usepackage{float}
\usepackage{cases}
\usepackage{mathtools}
\usepackage{dcolumn}
\usepackage{bbold}
\usepackage{bm}

\usepackage[dvipsnames]{xcolor}
\usepackage[colorlinks]{hyperref}
\hypersetup{
colorlinks=True,
linkbordercolor=White,
linkcolor=BrickRed,
citecolor=Blue,	 
}

\DeclarePairedDelimiter\bra{\langle}{\rvert}
\DeclarePairedDelimiter\ket{\lvert}{\rangle}

\renewcommand\vec[1]{\bm{#1}}

\newcommand\YMN[1]{#1}

\begin{document}

\title{Variability of electron and hole spin qubits due to interface roughness and charge traps}

\author{Biel Martinez}
\email{biel.martinezidiaz@cea.fr}
\affiliation{Univ. Grenoble Alpes, CEA, IRIG-MEM-L\_Sim, F-38000, Grenoble, France}
\author{Yann-Michel Niquet}
\email{yniquet@cea.fr}
\affiliation{Univ. Grenoble Alpes, CEA, IRIG-MEM-L\_Sim, F-38000, Grenoble, France}

\date{\today}

\begin{abstract}
Semiconductor spin qubits may show significant device-to-device variability in the presence of spin-orbit coupling mechanisms. Interface roughness, charge traps, layout or process inhomogeneities indeed shape the real space wave functions, hence the spin properties. It is, therefore, important to understand how reproducible the qubits can be, in order to assess strategies to cope with variability, and to set constraints on the quality of materials and fabrication. Here we model the variability of single qubit properties (Larmor and Rabi frequencies) due to disorder at the Si/SiO$_2$ interface (roughness, charge traps) in metal-oxide-semiconductor devices. We consider both electron qubits (with synthetic spin-orbit coupling fields created by micro-magnets) and hole qubits (with intrinsic spin-orbit coupling). We show that charge traps are much more limiting than interface roughness, and can scatter Rabi frequencies over one order of magnitude. We discuss the implications for the design of spin qubits and for the choice of materials.
\end{abstract}

\maketitle

\section{Introduction}

Spins in semiconductor quantum dots have emerged as a promising platform for quantum technologies.\cite{Loss98,Hanson07} Silicon\cite{Pla12,Zwanenburg13} and Germanium\cite{Scappucci20} are particularly attractive host materials because their main isotopes are free of nuclear spins that may interfere with the electron spins. Record spin lifetimes have thus been measured in isotopically purified $^{28}$Si samples.\cite{Tyryshkin12,Veldhorst14} High fidelity single and two qubit gates have been reported in a variety of silicon/silicon oxide and silicon/germanium devices,\cite{Kawakami14,Veldhorst15b,Kawakami16,Takeda16,Yoneda18,Watson18,Zajac18,Huang19,Xue19,Yang20,Petit20} and, more recently, a four qubit processor has been demonstrated in germanium (with hole spins).\cite{Hendrickx21}

The spin carrier can indeed be either an electron or a hole.\cite{Hendrickx21,Maurand16,Crippa18,Watzinger18,Hendrickx20,Hendrickx20b,Camenzind21} Electrons in the conduction band of silicon and germanium undergo much weaker spin-orbit coupling (SOC) than holes in the valence band.\cite{Winkler03,Kloeffel11,Kloeffel18} Therefore, electron spin qubits are better protected against electrical noise and undesirable interactions with, e.g., phonons. They tend to show, as a consequence, longer coherence and relaxation times than hole ``spin-orbit'' qubits.\cite{Li20,Lawrie20,CirianoTejel21} On the other hand, electron spin qubits can hardly be manipulated\cite{Corna18} electrically using Electric Dipole Spin Resonance (EDSR).\cite{Rashba03,Golovach06,Rashba08} To enable EDSR, an artificial SOC can be synthesized with micro-magnets that generate a gradient of magnetic field.\cite{Tokura06,Pioro-Ladriere07,Pioro-Ladriere08,Kawakami14,Kawakami16,Takeda16,Yoneda18,Watson18,Zajac18,Xue19,Yang20} The real space motion of the electron in this gradient indeed translates into an effective time-dependent magnetic field in the frame of the carrier, acting on its spin. Electrically driven spin rotations with Rabi frequencies up to a few MHz can be achieved that way in electron spin qubits (while tens of MHz are typically reported for holes with intrinsic SOC). Anyhow, introducing artificial SOC enhances the interactions of the spin with noise and phonons and tends to decrease the lifetimes of the qubits. The strength of SOC can be controlled by confinement and strains in hole spin-orbit qubits,\cite{Michal21} while it is set by the micromagnet design in electron spin qubits.

Another consequence of SOC for spin qubits is variability: the spin properties depend on the real space wave function, hence on inhomogeneities on a global scale (device-to-device variations in layout and characteristic sizes, wafer scale process fluctuations), and on a local scale (disorders such as interface roughness and charge traps). Such short range fluctuations are particularly annoying because they are random and cannot, therefore, be accounted for {\it a priori} in the design. As a consequence, each qubit must, in practice, be characterized individually before operating the quantum processor. While manageable for a few qubits, this may become a daunting bottleneck on the scale of hundreds of devices, especially if device-to-device variations are very large and cannot be at least partly compensated by bias adjustments or control software.

There is evidence in the literature that present spin qubits (with intrinsic or artificial SOC) show significant variability.\cite{Watson18,Yang20,Hendrickx20b,Camenzind21,Zwerver21} While the design of these devices is often responsible for part of this variability (non equivalent gate layouts for each qubit for example), it is important to understand how reproducible the qubits can be in the presence of uncontrolled local disorder, in order to set constraints on the quality of materials and fabrication.

The variability of metal-oxide-semiconductor (MOS) transistors has been extensively investigated and modeled,\cite{Bernstein06,Gareth06,Asenov09,Mezzomo11} yet systematic studies on semiconductor spin qubits are still scarce.\cite{Bourdet18,Ibberson18,Tong20,Simion20} In this work, we specifically model two sources of disorder at the Si/SiO$_2$ interface, interface roughness and charge traps (dangling bond $P_b$ defects), that are ubiquitous in MOS spin qubits. We focus on single qubit properties and discuss the distribution of the Zeeman splittings/Larmor frequencies and of the Rabi frequencies as a function of the strength of the disorder (height and lateral extent of the interface roughness fluctuations, density of traps). \YMN{The speed of one qubit gates in a quantum processor will indeed be limited by the smallest Rabi frequencies, while the overall lifetimes can be limited by the largest ones (since faster qubits are better coupled to the electric field and may, therefore, be more sensitive to noise)}. Also, large deviations of the individual Larmor and Rabi frequencies will complicate the management of radio-frequency control signals on the chip. We consider spin qubits in one-dimensional silicon-on-insulator channels as prototypical devices but most conclusions are generic and apply to other device layouts. We compare hole spin qubits with intrinsic SOC and electron spin qubits with artificial SOC and highlight the similarities and differences between the two kinds of carriers and SOC. We show that charge disorder in the vicinity of the qubits is much more critical than interface roughness, and can scatter Rabi frequencies over more than one order of magnitude. We finally discuss the implications for qubit lifetimes and performances, for device design and materials, and some solutions to mitigate variability.

The devices and methodology are introduced in section \ref{sec:Methodology}; the results for interface roughness are presented in section \ref{sec:Surfaceroughness}, and the results for charge traps in section \ref{sec:Chargetraps}. The implications and perspectives are discussed in section \ref{sec:Discussion}.

\section{Devices and methodology}
\label{sec:Methodology}

\begin{figure*}
\centering
\includegraphics[width=.85\columnwidth]{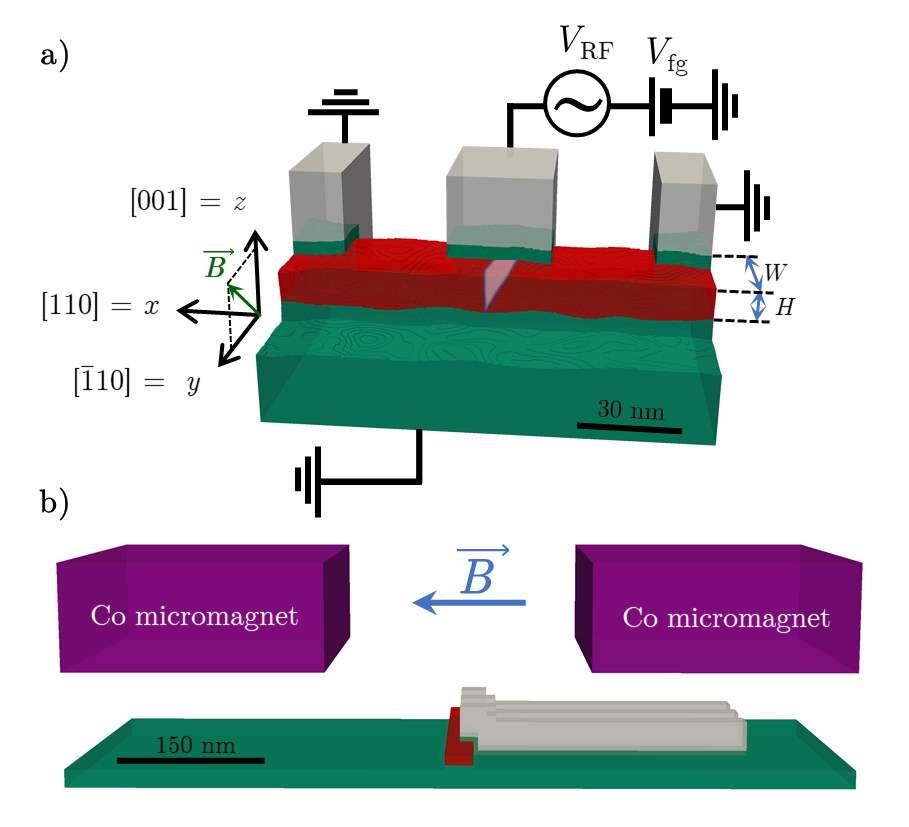}
\includegraphics[width=.95\columnwidth]{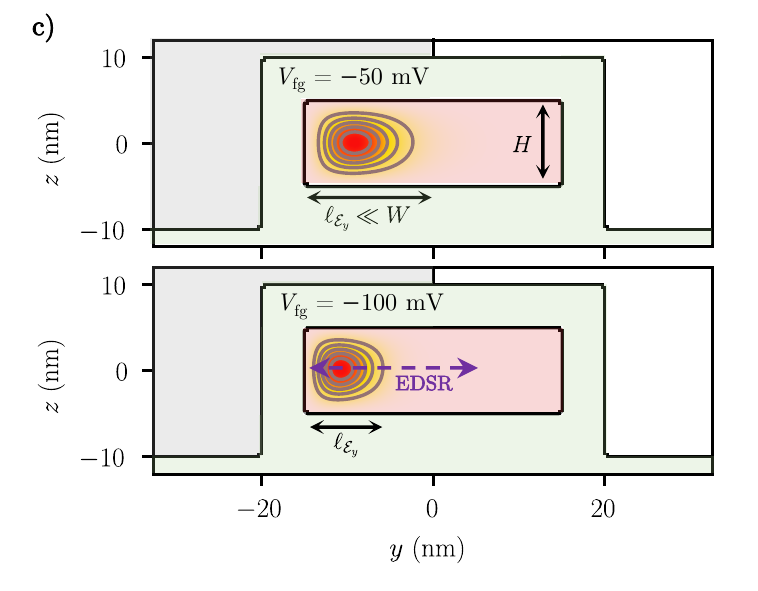}
\caption{(a) A sample of the rough devices generated for this study. The silicon channel is in red, the buried oxide in green and the gates in gray (the 5 nm thick gate oxide surrounding the whole channel has been partly removed for clarity). The dot is controlled by the central gate, while the substrate below the buried oxide and the side gates are grounded. The interface roughness is here characterized by rms $\Delta=0.4$ nm and correlation length $L_c=8$ nm. (b) Illustration of the micro-magnets layout used for electron devices. The Cobalt micro-magnets are in purple. The external magnetic field is oriented along (a) $\vec{x}+\vec{y}$ for holes, and (b) $\vec{y}$ for electrons (c) Maps of the squared ground-state hole wave function in the cross section of the channel at $V_{\rm fg}=-50$ mV and $V_{\rm fg}=-100$ mV. The cross-section plane is outlined by the shaded pink area in (a). \YMN{Vertical motion (along $z$) is dominated by structural confinement in the channel with thickness $H$, while lateral motion (along $y$) is dominated by electrical confinement by the top gate (with characteristic length scale $\ell_{\mathcal{E}_y}\ll W$). $\ell_{\mathcal{E}_y}$ decreases with increasingly negative $V_{\rm fg}$; therefore, a RF signal on the central gate (EDSR) drives the dot from left to right and modulates its lateral extension.}}
\label{fig:device}
\end{figure*}

In this section, we describe the devices and the methodology used in this study.

\subsection{Devices}

We have considered two classes of silicon devices: hole spin qubits with intrinsic SOC and electron spin qubits with a synthetic SOC created by micro-magnets. 

\subsubsection{Hole devices}

The hole spin qubits are similar to those of Refs. \onlinecite{Maurand16,Venitucci18,Li20} and are shown in Fig. \ref{fig:device}a. The devices are made of a $[110]$-oriented silicon nanowire channel with width $W=30$ nm [$(1\bar{1}0)$ facets] and height $H=10$ nm [$(001)$ facets] on top of a 25 nm thick buried silicon oxide and a silicon substrate. A 30 nm long central gate overlapping half of the nanowire controls a quantum dot. It is insulated from the channel by 5 nm of SiO$_2$. Two other gates on the left and right mimic neighboring qubits. The whole device is embedded in Si$_3$N$_4$. The silicon substrate can be used as a back gate, grounded throughout this study. The left and right gates are also grounded, and the depth of the dot potential is controlled by the central gate voltage $V_{\rm fg}$.

\YMN{Poisson’s equation for the single-particle potential in the empty dots} is solved with a finite volumes method (assuming dielectric constants $\kappa_{\rm Si}=11.7$, $\kappa_{{\rm SiO}_2}=3.9$, and $\kappa_{{\rm Si}_3{\rm N}_4}=7.5$). The ground-state Kramers pair is then computed with a finite-differences 6 bands $\vec{k}\cdot\vec{p}$ model\cite{Venitucci18} (assuming Luttinger parameters $\gamma_1=4.285$, $\gamma_2=0.339$, $\gamma_3=1.446$, split-off energy $\Delta=44$ meV, and Zeeman parameter $\kappa=-0.42$).\footnote{The valence band edge energy is set to $E_v=0$ eV in hole qubits; likewise, the conduction edge energy is set to $E_c=0$ eV in electron qubits.} This Kramers pair splits at finite magnetic field $B$ and can be manipulated electrically with a radio-frequency (RF) signal on the central gate, resonant with the Zeeman splitting $E_Z$ between the ``$\ket{\Uparrow}$'' and ``$\ket{\Downarrow}$'' (pseudo-)spin states. The single qubit can hence be characterized by the Larmor frequency $f_L=E_Z/h$, and by the Rabi frequency $f_R$, which quantifies the speed of the pseudo-spin rotations under electrical driving. Both are computed numerically with the $g$-matrix formalism of Ref. \onlinecite{Venitucci18}. This device operates in the ``$g$-tensor magnetic resonance'' ($g$-TMR) mode,\cite{Kato03} where the RF electric field from the central gate essentially modulates the \YMN{lateral confinement (see Fig. \ref{fig:device}c)}, hence the principal $g$-factors of the dot owing to the strong, intrinsic spin-orbit coupling in the valence bands.\cite{Venitucci18,Venitucci19,Michal21} We have verified that similar conclusions are achieved when the hole is driven \YMN{as a whole} along the channel by a RF signal on the left and right gates (``iso-Zeeman EDSR mode\cite{Michal21}'', see Appendix \ref{app:Additions}). 

The Larmor and Rabi frequencies of hole qubits are strongly dependent on the orientation of the external magnetic field $\vec{B}_{\rm ext}$. They have been computed for $\vec{B}_{\rm ext}$ along $\vec{y}+\vec{z}$, which is near the optimal direction in the $g$-TMR mode (see axes in Fig. \ref{fig:device}).\cite{Venitucci18,Venitucci19,Michal21} $f_L$ is proportional to the amplitude $B_{\rm ext}$ of the magnetic field, and $f_R$ to both $B_{\rm ext}$ and the amplitude $V_{\rm RF}$ of the driving signal $\delta V_{\rm fg}(t)=V_{\rm RF}\sin(2\pi f_L t)$ on the central gate. Therefore, $f_L$ and $f_R$ are normalized to reference $B_{\rm ext}=1$ T and $V_{\rm RF}=1$ mV throughout this work.

\subsubsection{Electron devices}

Intrinsic spin-orbit coupling is inefficient in the conduction band of silicon owing, in part, to the indirect nature of the band gap.\cite{Corna18} Therefore, a synthetic spin-orbit coupling must be introduced in order to allow for the electrical manipulation of electron spin qubits. This can be achieved by placing the qubits in a gradient of magnetic field provided by micro-magnets.\cite{Pioro-Ladriere08,Kawakami14,Takeda16,Yoneda18,Kawakami16} The RF electric field from the gate shakes the electron in this gradient; this translates into a time-dependent magnetic field in the frame of the electron, which acts on its spin. We add, therefore, two semi-infinite Co micro-magnets $100$ nm above the channel. They are $300$ nm thick and are split apart by $200$ nm (Fig. \ref{fig:device}b). We assume that the magnetic polarization $J=1.84$ T of the Co magnets\cite{Neumann15} is saturated in a transverse, external magnetic field $B_{\rm ext}=1$ T along $\vec{y}$. The vector potential created by these micro-magnets is calculated with a technique similar to Refs. \onlinecite{Yang90} and \onlinecite{Goldman00} (see Appendix \ref{app:micromagnets}); and its action on the real space and spin motions is described by an anisotropic effective mass model for the Z-valleys (with longitudinal mass $m_l^*=0.916\,m_0$ and transverse mass $m_t^*=0.191\,m_0$, see Appendix \ref{app:Rabi}). We discard spin-valley and valley-orbit interactions in this study. The effects of interface roughness/steps and charge traps on the valley splitting have been discussed, for example, in Refs. \onlinecite{Culcer10,Jiang12,Bourdet18,Ibberson18,Abadillo18,Tariq19,Hosseinkhani20} \YMN{(see further discussion in Section \ref{sec:Sensitivity})}.

\subsection{Disorders and methodology}

In this work, we focus on two kinds of disorder at the Si/SiO$_2$ interface: interface roughness and charge traps ($P_b$ defects). \YMN{We also briefly address disorder in the micro-magnets in Appendix \ref{app:Varmicro}.}

\subsubsection{Interface roughness}

Interface roughness is explicitly added in the solvers for Poisson's equation and for the 6 bands $\vec{k}\cdot\vec{p}$ or effective mass models. For that purpose, samples of roughness are randomly generated on each facet of the wire with a target Gaussian auto-correlation function:\cite{Goodnick85}
\begin{equation}
E[\delta h(\vec{R}_\parallel) \delta h(\vec{R}_\parallel+\vec{r}_\parallel)]=\Delta^2 e^{-r_\parallel^2/L_c^2}\,,
\label{eq:autoSR}
\end{equation}
where $\delta h(\vec{r}_\parallel)$ is the out-of-plane displacement of the interface, $\vec{R}_\parallel$ and $\vec{r}_\parallel$ are in-plane positions, and $E[\cdot]$ denotes an ensemble average. $\Delta$ is the rms amplitude of the roughness, and $L_c$ is a correlation length that characterizes the lateral extent of the fluctuations. The samples are generated in Fourier space, then transformed to real space, along the lines of Ref. \onlinecite{Niquet14}.

In micro-electronics grade devices, atomic scale TEM images and room-temperature mobility measurements suggest $\Delta$ in the $0.2-0.4$ nm range and $L_c$ in the 1 to 4 nm range.\cite{Goodnick85,Pirovano00,Esseni03,Nguyen14,Bourdet16,Zeng17b} However, the interface roughness limited mobility is typically measured at high carrier density, and therefore probes short length scales fluctuations, whereas spin qubits operate at low carrier density, and are thus presumably more sensitive to longer length scale fluctuations (as revealed, for example, by atomic force microscopy\cite{Pirovano00}). We have, therefore, varied $L_c$ between 1 and 30 nm\YMN{, in order to span the experimentally relevant range and allow for a systematic exploration of the trends}.

\subsubsection{Charge traps}

The Si/SiO$_2$ interface separates a crystalline and an amorphous material, and is therefore prone to be defective. The prototype of these defects is the dangling bond ($P_b$) at the interface.\cite{Poindexter84,Gerardi86,Poindexter89,Helms94,Thoan11} $P_b$ defects are amphoteric: they trap electrons on shallow to deep acceptor levels in $n$-type devices, and trap holes on donor levels in $p$-type devices. For simplicity, we model the charged defects as negative ($-e$) point charges at the interface in electron spin qubits, and as positive ($+e$) point charges in hole spin qubits, with densities $n_i$ ranging from $10^{10}$ to $10^{11}$ cm$^{-2}$. These densities are, again, typical of micro-electronics grade interfaces.\cite{Bauza02,Brunet09,Pirro16,Vermeer21} Our point charges model neglects the finite extension of the bound charge density around the defects; however, this is little relevant as the majority carriers in the dots get repelled by a charged $P_b$, and, therefore, do not probe much the Coulomb singularity.

\subsubsection{Methodology}

The Larmor and Rabi frequencies are collected on sets on $N\ge 500$ random samples of disorder (either interface roughness or charge traps).\YMN{\footnote{\YMN{We model one device at a time, with a different seed for the random number generator used by the geometry builder.}}} Then the average Larmor or Rabi frequency $\overline{f}$, and its standard deviation $\sigma(f)$ are estimated as:\footnote{\YMN{The factor $N-1$ (instead of $N$) on the denominator of Eq. (\ref{eq:sigma}) follows from the definition of the ``unbiased'' estimate of the variance. This so-called Bessel correction does not make significant differences given the large data sets considered here ($N>500$).}}
\begin{subequations}
\begin{align}
\overline{f}&=\frac{1}{N}\sum_{i=1}^N f_i \\
\sigma(f)&=\left[\frac{1}{N-1}\sum_{i=1}^N\left(f_i-\overline{f}\right)^2\right]^{1/2}\,, \label{eq:sigma}
\end{align}
\end{subequations}
where the $f_i$ are the sampled frequencies. We characterize the variability by the relative standard deviation (RSD):
\begin{equation}
\tilde{\sigma}(f)=\sigma(f)/\overline{f}\,,
\end{equation}
and by the inter-quartile range:
\begin{equation}
{\rm IQR}(f)=\hat{f}(0.75)-\hat{f}(0.25)\,,
\end{equation}
where $\hat{f}(\alpha)$ is the frequency such that a proportion $\alpha$ of the devices have $f_i<\hat{f}(\alpha)$. Therefore, 50\% of the qubits lie within the IQR. Confidence intervals on $\overline{f}$, $\tilde{\sigma}(f)$ and the IQR are estimated using a percentile bootstrap (resampling) method.\cite{Davison97}

\section{Interface roughness}
\label{sec:Surfaceroughness}

\subsection{Holes}

\begin{figure}
\centering
\includegraphics[width=.95\columnwidth]{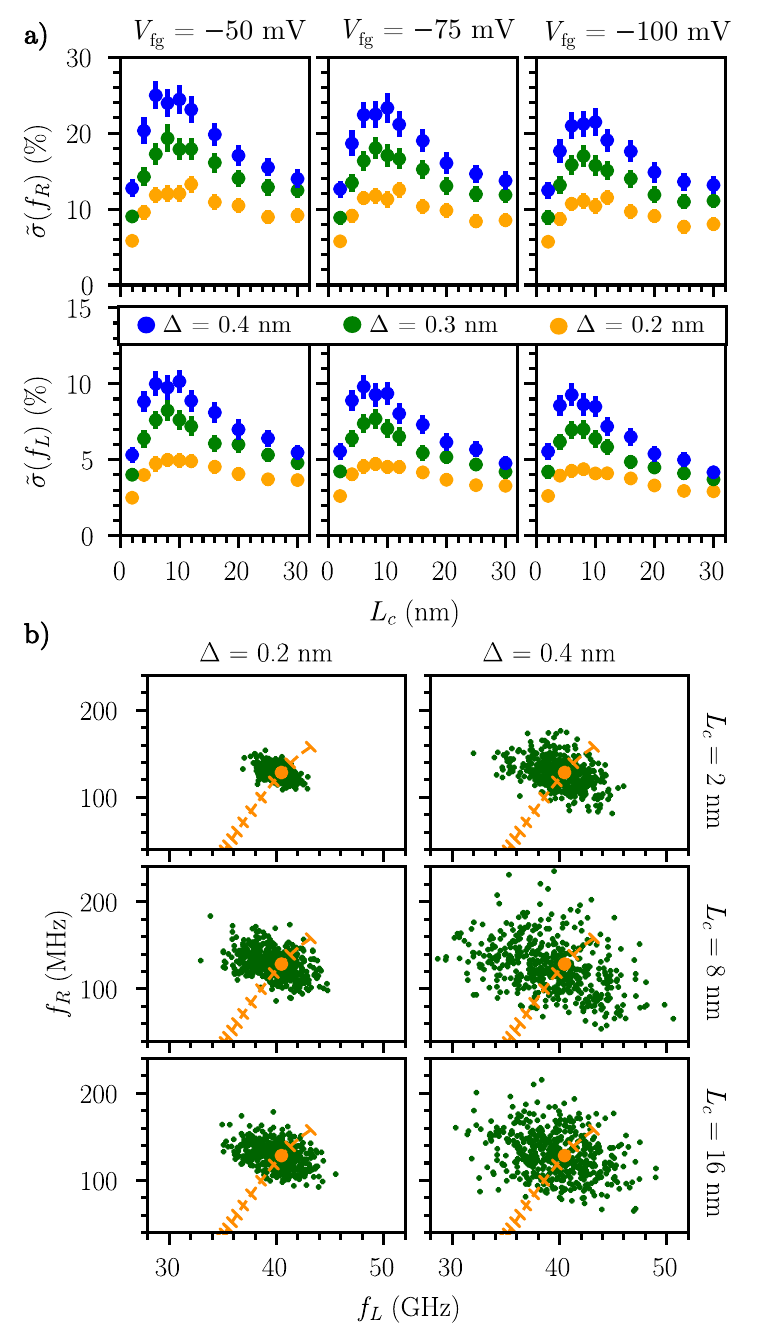}
\caption{(a) RSD $\tilde{\sigma}(f_R)$ of the Rabi frequency and RSD $\tilde{\sigma}(f_L)$ of the Larmor frequency of rough hole qubits as a function of $L_c$ for different $\Delta$ and gate voltages $V_{\rm fg}$. The error bars outline the 95\% confidence interval. (b) Distribution of the rough hole devices in the $(f_L, f_R)$ plane for different $\Delta$ and $L_c$ at $V_{\rm fg}=-50$ mV. Each green point is a particular realization of the interface roughness disorder. The orange dashed line is the pristine device frequencies as a function of $V_{\rm fg}$ (crosses by steps of 10 mV, increasingly negative from top right to down left, with the orange point at $V_{\rm fg}=-50$ mV).}
\label{fig:SRHoles}
\end{figure}

The RSD $\tilde{\sigma}(f_R)$ of the Rabi frequency and the RSD $\tilde{\sigma}(f_L)$ of the Larmor frequency of rough hole qubits are plotted as a function of $L_c$ for different $\Delta$ and front gate voltages $V_{\rm fg}$ in Fig. \ref{fig:SRHoles}a. They are normalized with respect to the average Rabi frequency $\overline{f_R}\simeq f_R^0$ and average Larmor frequency $\overline{f_L}\simeq f_L^0$, which are very close to the Rabi frequency $f_R^0$ and Larmor frequency $f_L^0$ of the pristine device given in Table \ref{tab:fLR0H}. The inter-quartile ranges are ${\rm IQR}(f_R)\simeq 4\sigma(f_R)/3$ and ${\rm IQR}(f_L)\simeq 4\sigma(f_L)/3$, as expected for quasi-normal distributions.

\begin{table}[t]
\centering
\begin{tabular}{r|r|r|r}
\toprule
$V_{\rm fg}$ (mV) & $f_L^0$ (GHz) & $f_R^0$ (MHz) & $\ell_y^0$ (nm) \\ 
\hline
$-50$ & 40.46 & 128.47 & 3.28 \\
$-75$ & 37.62 & 84.00 & 2.71\\
$-100$ & 36.06 & 56.19 & 2.38\\
\botrule
\end{tabular}
\caption{Larmor frequency $f_L^0$ and Rabi frequency $f_R^0$ of the pristine hole devices, computed at $B_{\rm ext}=1$ T along $\vec{y}+\vec{z}$, and $V_{\rm RF}=1$ mV. The confinement length $\ell_y^0=\sqrt{\langle y^2\rangle-\langle y\rangle^2}$ is also given (see section \ref{sec:Mechanisms}).}
\label{tab:fLR0H}
\end{table}

The interface roughness primarily modulates the thickness of the channel. This alters the shape of the hole envelopes, hence the dipole or momentum matrix elements relevant for the Larmor and Rabi frequencies.\cite{Michal21} This will be further analyzed in section \ref{sec:Mechanisms}.

On the one hand, the variability increases, as expected, with the rms amplitude $\Delta$ of the interface roughness fluctuations (almost linearly in this range). On the other hand, the variability shows a non-monotonic behavior with respect to the correlation length $L_c$ of the fluctuations, and peaks near $L_c=10$ nm. Indeed, fluctuations with length scales much shorter than the characteristic size of the dot tend to be averaged out and have little effect on the envelope functions of the holes. On the opposite, the interfaces become flat again on the scale of the dot when $L_c$ is much greater than the gate dimensions. Yet the variability does not tend to zero when $L_c\to\infty$: in fact, the thickness and width of the channel can still vary from device to device (with rms $\sqrt{2}\Delta$) as top/down, left/right interface fluctuations are uncorrelated. The variability is maximal when the length scale of the fluctuations is comparable with the size of the dot (see section \ref{sec:Mechanisms}).

As the front gate voltage is made further negative, the dot gets squeezed on the side of the channel by the lateral electric field $\mathcal{E}_y$ (see Fig. \ref{fig:device}c and Table \ref{tab:fLR0H}).\cite{Venitucci18,Venitucci19} In this regime, the electric confinement length shall prevail over the characteristic scales of disorder on the top and bottom facets of the channel. The absolute standard deviations $\sigma(f_L)$ and $\sigma(f_R)$, are, therefore, expected to decrease. However, $f_R^0$ also decreases continuously with increasingly negative $V_{\rm fg}$ because the strongly confined dot can hardly be shaken anymore by the RF electric field.\cite{Venitucci19,Michal21} As a consequence, the RSD $\tilde{\sigma}(f_R)$ shows a rather weak scaling with $V_{\rm fg}$. The RSD $\tilde{\sigma}(f_L)$ exhibits an even weaker scaling even though the Larmor frequency $f_L^0$ remains finite at large negative $V_{\rm fg}$. As discussed in \ref{sec:Mechanisms}, this results from the increasing role of the disorder on the side facets.

For a rms $\Delta=0.3$ nm, $\tilde{\sigma}(f_L)$ can be as large as  $8\%$, and $\tilde{\sigma}(f_R)$ as large as $20\%$ ($V_{\rm fg}=-50$ mV). The distribution of devices in the $(f_L, f_R)$ plane is plotted in Fig. \ref{fig:SRHoles}b. There are no clear correlations between the deviations of Larmor and Rabi frequencies. The $(f_L(V_{\rm fg}), f_R(V_{\rm fg}))$ line of the pristine device is also reported on this figure. It outlines the average electrical tunability of the Larmor and Rabi frequencies. The deviations of Larmor frequencies must, in particular, be correctable if one wishes to address all qubits at a well defined RF frequency. The extent of bias corrections is, however, limited by the stability diagram of the device. Assuming a charging energy $U\simeq 10$ meV typical for these qubits, the range of gate voltage giving rise to chemical potential shifts $|\delta\mu|<U/2$ in the dot is only $|\delta V_{\rm fg}|\lesssim 10$ meV. This limitation can nonetheless be overcome if inter-dot tunneling can be cut down sufficiently during single qubit operations. The deviations of Rabi frequencies may also be corrected by tuning the driving RF power for each qubit. This will be further discussed in section \ref{sec:Discussion}. 

\subsection{Electrons}

\begin{figure}
\centering
\includegraphics[width=.95\columnwidth]{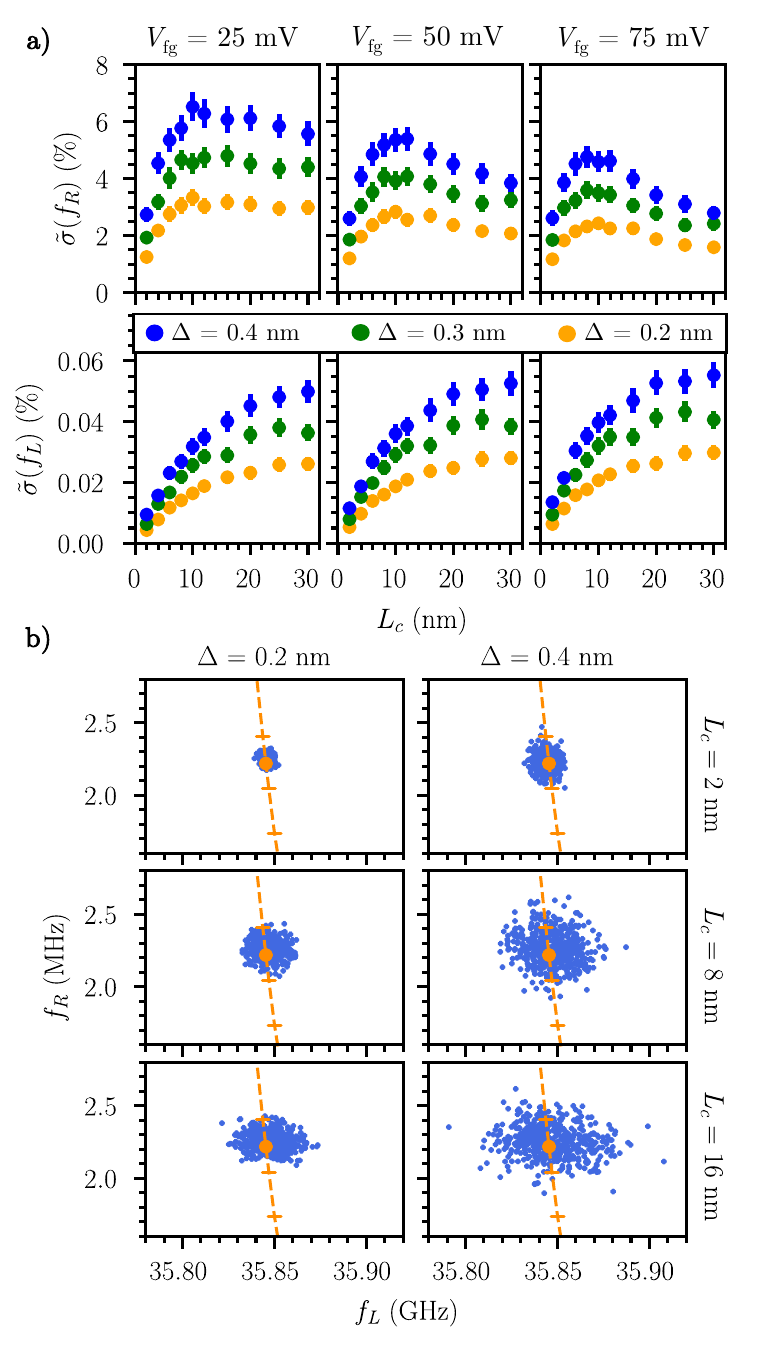}
\caption{(a) RSD $\tilde{\sigma}(f_R)$ of the Rabi frequency and RSD $\tilde{\sigma}(f_L)$ of the Larmor frequency of rough electron qubits as a function of $L_c$ for different $\Delta$ and gate voltages $V_{\rm fg}$. They are normalized with respect to the average Rabi frequency $\overline{f_R}\simeq f_R^0$ and average Larmor frequency $\overline{f_L}\simeq f_L^0$ given in Table \ref{tab:fLR0E}. The error bars outline the 95\% confidence interval. (b) Distribution of the rough electron devices in the $(f_L, f_R)$ plane for different $\Delta$ and $L_c$ at $V_{\rm fg}=50$ mV. Each blue point is a particular realization of the interface roughness disorder. The orange dashed line is the pristine device frequencies as a function of $V_{\rm fg}$ (crosses by steps of 10 mV, increasing from top right to down left, with the orange point at $V_{\rm fg}=50$ mV).}
\label{fig:SRElectrons}
\end{figure}

The data for electron qubits are shown in Fig. \ref{fig:SRElectrons}.

\begin{table}[t]
\centering
\begin{tabular}{r|r|r|r}
\toprule
$V_{\rm fg}$ (mV) & $f_L^0$ (GHz) & $f_R^0$ (MHz) & $\ell_y^0$ (nm) \\ 
\hline
$25$ & 35.837 & 3.20 & 5.20 \\
$50$ & 35.845 & 2.22 & 4.69 \\
$75$ & 35.853 & 1.48 & 4.21\\
\botrule
\end{tabular}
\caption{Larmor frequency $f_L^0$ and Rabi frequency $f_R^0$ of the pristine electron devices, computed at $B_{\rm ext}=1$ T along $\vec{y}$ and $V_{\rm RF}=1$ mV. The confinement length $\ell_y^0=\sqrt{\langle y^2\rangle-\langle y\rangle^2}$ is also given (see section \ref{sec:Mechanisms}).}
\label{tab:fLR0E}
\end{table}

First of all, the Rabi frequencies of the pristine electron qubits (Table \ref{tab:fLR0E}, in the MHz range) are much smaller than the Rabi frequencies of the pristine hole qubits (Table \ref{tab:fLR0H}, in the tens of MHz range). These ranges are consistent with the Rabi frequencies measured in electron and hole spin qubits with different layouts.\cite{Kawakami16,Watson18,Yang20,Maurand16,Watzinger18,Hendrickx20b} Intrinsic spin-orbit coupling in the valence band is indeed, much more efficient than synthetic spin-orbit coupling in the conduction band. As a consequence, fast electrical driving calls for much larger real space motion (larger electric dipoles) in electron than in hole qubits, hence for larger/shallower dots. For the sake of consistency, we nonetheless compare electron and hole qubits with the same topology and structural dimensions here. \YMN{Alternatively, the Rabi frequencies may be increased using stronger micro-magnets or bringing them closer to the qubits, when technologically feasible.}

The variability of Larmor frequencies is as low as $0.05\%$, but increases monotonously and saturates at large $L_c$. It results from the vertical displacement of the dot as a whole in the gradient of magnetic field, induced by the interface roughness (see discussion in the next section).

The trends for the Rabi frequencies are the same as in hole qubits. The dependence of $\tilde{\sigma}(f_R)$ on gate voltage is slightly stronger for electrons than for holes; yet variability remains much smaller in electron qubits ($\tilde{\sigma}(f_R)<5\%$ at $\Delta=0.3$ nm and $V_{\rm fg}=50$ mV). We emphasize, though, that electron qubits are  subject to additional sources of variability \YMN{(e.g., alignment and homogeneity of the micro-magnets\cite{Yoneda15,Simion20}) that will be discussed in section \ref{sec:Sensitivity}}.

We analyse below the mechanisms responsible for variability and the differences between electron and hole qubits.

\subsection{Mechanisms responsible for the variability}
\label{sec:Mechanisms}

\begin{figure}
\centering
\includegraphics[width=.95\columnwidth]{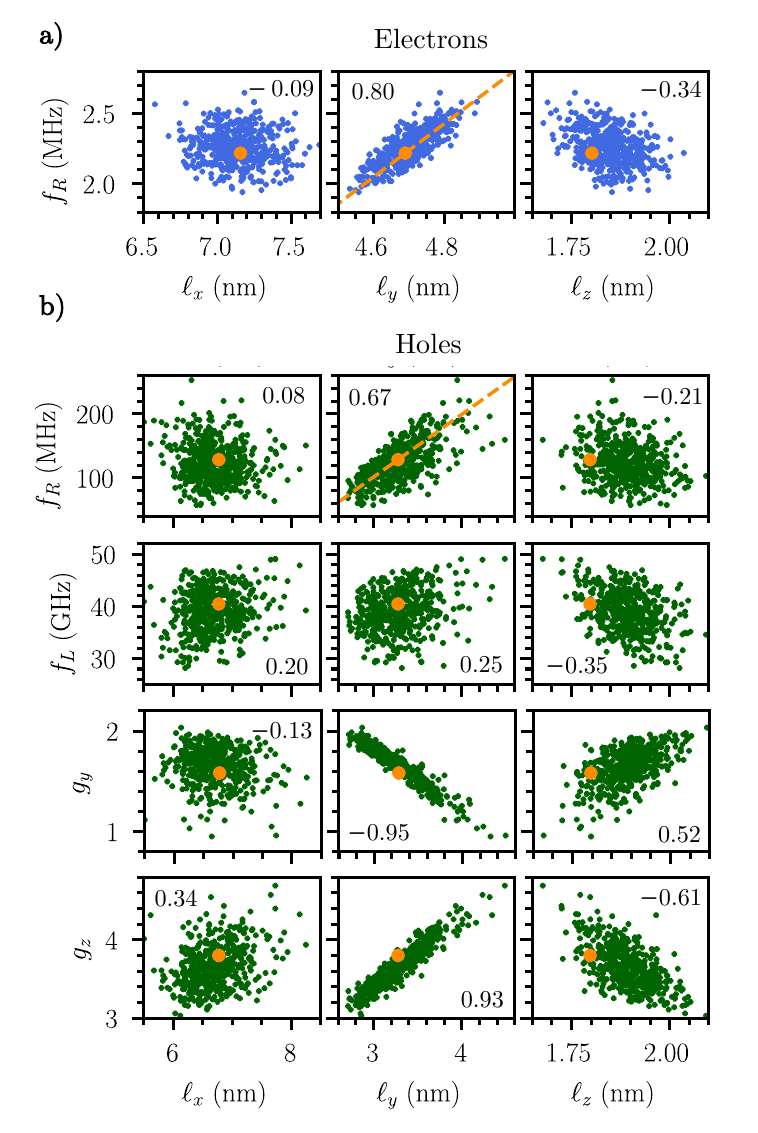}
\caption{(a) Correlations between the Rabi frequency $f_R$ of the electron qubits and the extension $\ell_x$, $\ell_y$ and $\ell_z$ of the dots along the $x$, $y$, and $z$ axes respectively. Each point is a particular realization of interface roughness disorder with $\Delta=0.4$ nm and $L_c=10$ nm ($V_{\rm fg}=50$ mV). The orange dot is the pristine device. \YMN{The correlation coefficient $\rho$ [Eq. (\ref{eq:corr})] between the abscissas and ordinates is given in each panel.} The dashed orange line $\delta f_R/f_R^0=4\delta\ell_y/\ell_y^0$ is a guide to the eye for the discussion of section \ref{sec:Mechanisms}. (b) Same for hole qubits: correlations between the Rabi frequency $f_R$, the Larmor frequency $f_L$, the gyromagnetic factors $g_y$ and $g_z$, and the extension of the dots $\ell_x$, $\ell_y$ and $\ell_z$ ($V_{\rm fg}=-50$ mV). The dashed orange line $\delta f_R/f_R^0=3\delta\ell_y/\ell_y^0$ is a guide to the eye.}
\label{fig:SRcorrelations}
\end{figure}

The electrons and holes are confined in the cross section of the channel by the vertical component $\mathcal{E}_z$ and the lateral component $\mathcal{E}_y$ of the electric field from the central gate. The strength of this electric field can, therefore, be characterized by the electric confinement lengths $\ell_{\mathcal{E}_z}=[\hbar^2/(2m_\perp^*e\mathcal{E}_z)]^{1/3}$ and $\ell_{\mathcal{E}_y}=[\hbar^2/(2m_\parallel^*e\mathcal{E}_y)]^{1/3}$, where $m_\perp^*$ is the vertical confinement mass along $[001]$, and $m_\parallel^*$ the in-plane mass.\cite{Michal21} In the regimes explored in this work, $\ell_{\mathcal{E}_z}\gtrsim H$ while $\ell_{\mathcal{E}_y}\lesssim W$: the vertical confinement is dominated by the structure, while the in-plane confinement is dominated by the electric field \YMN{(see Fig. \ref{fig:device}c)}.

When $H\ll\ell_{\mathcal{E}_z}$, the interface roughness on the main top and bottom facets essentially modulates the thickness $H$ of the channel and the associated structural confinement energy $E_\perp=\hbar^2\pi^2/(2m_\perp^*H^2)$. Therefore, in a single band model, long wavelength thickness fluctuations of $\delta H(x, y)$ translate into a potential
\begin{equation}
W(x,y)\approx \delta H(x,y)\frac{\partial E_\perp}{\partial H}\approx -\delta H(x,y)\frac{\hbar^2\pi^2}{m_\perp^*H^3}
\label{eq:WSR}
\end{equation}
for the motion in the weakly confined $(xy)$ plane.\cite{Sakaki87,Uchida03,Jin07} In the opposite limit $\ell_{\mathcal{E}_z}\ll H$, $W(x,y)\approx -e\mathcal{E}_z\delta z_t(x,y)$ would be dominated by the fluctuations of the position $z_t(x,y)$ of the top interface in the vertical electric field. This regime is, however, more relevant for the iso-Zeeman driving mode of the holes (whose optimum is at strong field), and for the contribution of the side facets (whose role is discussed below).

The dot essentially oscillates along $y$ when the driving RF signal is applied to the central gate because the electric dipole is small along the strong confinement axis $z$. The model for the electron qubits developed in Appendix \ref{app:Rabi} shows that the Rabi frequency of the electrons is then proportional to $(\partial B_z/\partial y)\times(\partial\langle y\rangle/\partial V_{\rm fg})$, where $\langle y\rangle=\bra{\psi}y\ket{\psi}$ is the expectation value of the in-plane $y$ coordinate in the ground-state $\ket{\psi}$ at zero magnetic field. The interface roughness shapes the envelope function of the electron in the $(xy)$ plane hence the electrical response $\partial\langle y\rangle/\partial V_{\rm fg}$ of the dot.

In the simplest models for the dot (hard-wall or harmonic confinement potential with homogeneous electric field along $y$),\cite{Venitucci19, Michal21} a dimensional analysis of the perturbation series for $\langle y\rangle(V_g)$ suggests that $\partial\langle y\rangle/\partial V_{\rm fg}\propto\ell_y^4$, where $\ell_y=\sqrt{\langle y^2\rangle-\langle y\rangle^2}$ is the extension of the dot along $y$: the stronger the confinement, the weaker the electrical response. Although the disordered potential considered here is rather complex, we can still expect correlations between the lateral size of the dot and the Rabi frequency.

In order to highlight such correlations, we have plotted the Rabi frequency $f_R$ of the electron qubits as a function of the extensions $\ell_x=\sqrt{\langle x^2\rangle-\langle x\rangle^2}$, $\ell_y$, and $\ell_z=\sqrt{\langle z^2\rangle-\langle z\rangle^2}$ of the ground-state in Fig. \ref{fig:SRcorrelations}a ($V_{\rm fg}=50$ mV, $\Delta=0.4$ nm, $L_c=10$ nm). \YMN{The correlation coefficient $\rho$ between the abscissas ($X=\ell_x,\,\ell_y$ or $\ell_z$) and the ordinates ($Y=f_R$), defined as:
\begin{align}
\rho&=\frac{1}{\sigma(X)\sigma(Y)}E\left[(X-\overline{X})(Y-\overline{Y})\right] \nonumber \\
&\equiv\frac{1}{\sigma(X)\sigma(Y)(N-1)}\sum_{i=1}^{N}(X_i-\overline{X})(Y_i-\overline{Y})\,,
\label{eq:corr}
\end{align}
is reported in each panel ($\rho=0$ for uncorrelated and $|\rho|\to 1$ for linearly correlated $X$ and $Y$). The correlations are much more significant between $f_R$ and $\ell_y$ than between $f_R$ and $\ell_x$ or $\ell_z$}, and follow approximately the relation $\delta f_R/f_R^0\approx 4\delta\ell_y/\ell_y^0$ expected from the $\propto\ell_y^4$ dependence of $\partial\langle y\rangle/\partial V_{\rm fg}$ (with $\delta f_R=f_R-f_R^0$, $\delta\ell_y=\ell_y-\ell_y^0$, and $\ell_y^0$ the extension of the ground-state in the pristine device). This supports the conclusion that the Rabi frequency is primarily modulated by the distortions of the dot along $y$ induced by the interface roughness. Note that the distribution of $\ell_z$ is skewed towards large $\ell_z>\ell_z^0$, as the electron tends to localize in the thickest parts of the channel.\footnote{As a matter of fact, $\ell_z^0=\sqrt{\pi^2-6}H/(2\sqrt{3}\pi)\approx0.18H$ in a thin film with thickness $H$. Since 95\% of the thickness fluctuations lie within a $\pm3\Delta$ range, and the electron/hole tends to localize in the thickest parts of the channel ($\Delta>0$), we expect $0\lesssim\ell_z-\ell_z^0\lesssim0.18\times 3\Delta\approx 0.22$ nm, in agreement with Fig. \ref{fig:SRcorrelations}}

\begin{figure}
\centering
\includegraphics[width=.95\columnwidth]{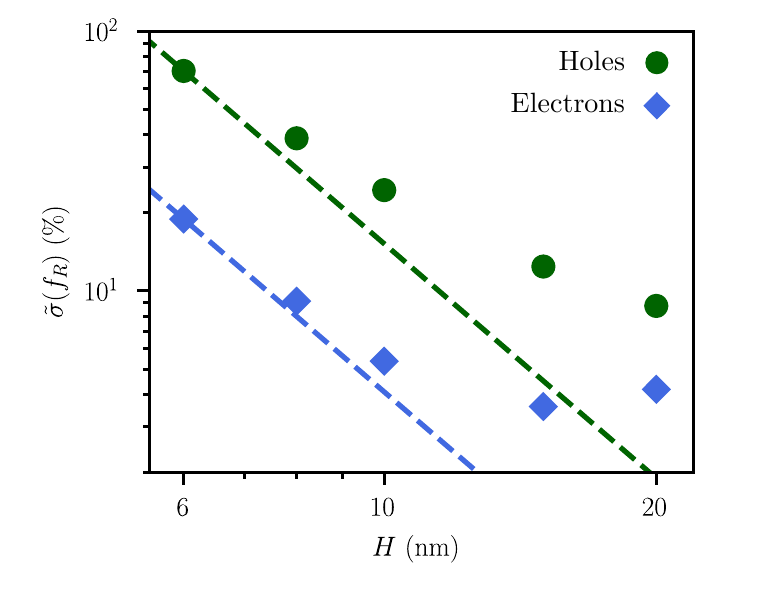}
\caption{Dependence of the RSD $\tilde{\sigma}(f_R)$ of the Rabi frequency of electrons and hole qubits on the channel thickness $H$ (interface roughness disorder with $\Delta=0.4$ nm and $L_c=10$ nm; $V_{\rm fg}=+50$ mV for electrons and $V_{\rm fg}=-50$ mV for holes). \YMN{The dashed lines are $\propto 1/H^3$ extrapolations from $H=6$ nm}.}
\label{fig:Thickness}
\end{figure}

A simple analytical model for variability in a parabolic in-plane confinement potential, based on first-order perturbation theory, is discussed in Appendix \ref{app:Model}. This model explains most trends outlined above, even though confinement is highly anharmonic in the present devices. It confirms, in particular, that $\tilde{\sigma}(f_R)\propto\Delta/(m_\perp^*H^3)$ in rough films with thickness $H\ll\ell_{\mathcal{E}_z}$, as evidenced by Fig. \ref{fig:Thickness}. It also highlights that $\tilde{\sigma}(f_R)$ increases, in general, with the in-plane mass $m_\parallel^*$: the heavier the particles, the stronger they localize in the disorder. Finally, this model shows that the variability peaks when $L_c\approx 2\ell_y^0$: fluctuations with wavelengths comparable to the size of the dot are expected to have the largest impact on its shape. The data of Fig. \ref{fig:SRElectrons} peak at a slightly larger $L_c$, owing, likely, to the anharmonicity of the in-plane confinement.

The analysis is more intricate for holes. The model of Ref. \onlinecite{Michal21} suggests that the Rabi frequency of the holes is inversely proportional to the gap $\Delta_{\rm LH}\propto 1/H^2$ between the heavy- and light-hole sub-bands, and proportional to $\ell_y^\alpha$, where $\alpha$ can range from 1 ($W\ll\ell_{\mathcal{E}_y}$) to 4 ($W\gg\ell_{\mathcal{E}_y}$). This behavior essentially results from the fact that $f_R\propto\partial\langle k_y^2\rangle/\partial V_{\rm fg}$ instead of $\propto\partial\langle y\rangle/\partial V_{\rm fg}$, where $k_y=-i\partial/\partial y$ and the expectation value is computed for the ground-state heavy-hole envelope function. We may, therefore, expect correlations between $f_R$ and $\ell_y$ and/or $\ell_z$. Figure \ref{fig:SRcorrelations}b actually shows \YMN{stronger correlations with $\ell_y$ ($\delta f_R/f_R^0\approx 3\delta\ell_y/\ell_y^0$) than with $\ell_z$}. The variability of the Rabi frequency is still, therefore, dominated by fluctuations of the in-plane size of the dot (at least for $L_c<20$ nm). 

As for the Larmor frequency of electrons, $\delta f_L=f_L-f_L^0$ depends on the position of the dot in the inhomogeneous field of the micro-magnets. The largest components of the gradient are $\partial B_z/\partial y=\partial B_y/\partial z$ (see Appendix \ref{app:micromagnets}), but only the latter gives rise to first-order variations of the Larmor frequency in a static magnetic field along $\vec{y}$ (while the former is responsible for the Rabi oscillations). As a consequence, the Larmor frequency correlates with $\langle z\rangle$ instead of $\ell_y$. The spread of $\langle z\rangle$ is expected to grow monotonously with increasing $L_c$; when $L_c\to\infty$, the roughness profile becomes flat again on the scale of the dot so that
\begin{equation}
\langle z\rangle\approx\frac{1}{2}\left(z_b+z_t\right)\,,
\end{equation}
where $z_b$ and $z_t$ are the positions of the bottom and top interfaces. As $\sigma((z_b+z_t)/2)=\Delta/\sqrt{2}$ (the interfaces being uncorrelated), 
\begin{equation}
\tilde{\sigma}(f_L)\to\frac{1}{B_y}\frac{\partial B_y}{\partial z}\frac{\Delta}{\sqrt{2}}\,,
\end{equation}
where $B_y$ is the total (external and micro-magnets) field along $y$ at the center of the channel. In the present layout, $\tilde{\sigma}(f_L)$ remains small because the inhomogeneous component of the magnetic field $\Delta\times\partial B_y/\partial z$ is much lower than the external magnetic field $B_{\rm ext}=1$ T. 

For holes, the heavy-hole ground state gets mixed with a light-hole component by the electric field $\mathcal{E}_y$,\cite{Venitucci18} due to the competition between vertical (\YMN{mostly} structural) confinement along $z$ and lateral (electric) confinement along $y$. As a consequence, the gyromagnetic factor $g_z$ decreases while $g_y$ increases with decreasing $\ell_y$ \YMN{and increasing $\ell_z$} (see Figure \ref{fig:SRcorrelations}b).\cite{Michal21} \YMN{Both $g_y$ and $g_z$ show in fact stronger correlations with $\ell_y$ than with $\ell_z$ because the system is more polarizable along $y$ than along $z$. The Larmor frequency $f_L\simeq\mu_BB_{\rm ext}\sqrt{(g_y^2+g_z^2)/2}$ is, therefore, also expected to show dominant correlations with $\ell_y$, even though the variations of $g_y$ and $g_z$ do partly cancel}. The correlations between $f_L$ and $\ell_y$ are, actually, further blurred by the fact that $y$ and $z$ are not principal axes of the $g$-tensor anymore in rough channels, as the coupling between in and out-of-plane motions introduces an extra off-diagonal $g_{yz}$ factor in the above expression for $f_L$.\footnote{In the pristine device, $x$, $y$ and $z$ are very good approximations to the principal axes of the $g$-matrix.\cite{Venitucci18} In disordered devices, the channel axis $x$ remains, in general, a good principal axis; yet in the presence of interface roughness (but not charge traps), the two other principal axes can make an angle of up to $\approx 10$ degrees with $y$ and $z$. This rotation results from the coupling between the in-plane and out-of-plane motions induced by the interface roughness. Although small, it has sizable effects on the distribution of Larmor frequencies.} Finally, the balance between vertical and lateral confinement is also controlled by interface roughness on the lateral facets of the channel, which play an increasing role when decreasing $V_{\rm fg}<0$; this partly explains why $\tilde{\sigma}(f_L)$ (and to a lesser extent $\tilde{\sigma}(f_R)$) shows little dependence on $V_{\rm fg}$ at least in the investigated bias range. 

The variability due to interface roughness is significantly larger in hole than in electron qubits (also see Fig. \ref{fig:Thickness}). For the Larmor frequencies, this results from the dependence of the $g$-factors of holes on the confinement discussed above. For the Rabi frequencies, this mostly results from the different electron and hole effective masses. Indeed, the in-plane mass of electrons and holes are similar ($m_\parallel^*\simeq 0.2\,m_0$), but the confinement mass $m_\perp^*$ are different: $m_\perp^*=m_l^*=0.916\,m_0$ for electrons and $m_\perp^*=m_0/(\gamma_1-2\gamma_2)=0.277\,m_0$ for heavy holes. According to Eq. (\ref{eq:WSR}), the interface roughness has, therefore, stronger impact on the hole than on the electron qubits when $H\ll\ell_{\mathcal{E}_z}$ (vertical confinement dominated by the structure). This may not be the case, however, in thicker channels operating in the $\ell_{\mathcal{E}_z}\ll H$ limit (vertical confinement dominated by the electric field).

\section{Charge traps}
\label{sec:Chargetraps}

\subsection{Holes}

\begin{figure}
\centering
\includegraphics[width=.95\columnwidth]{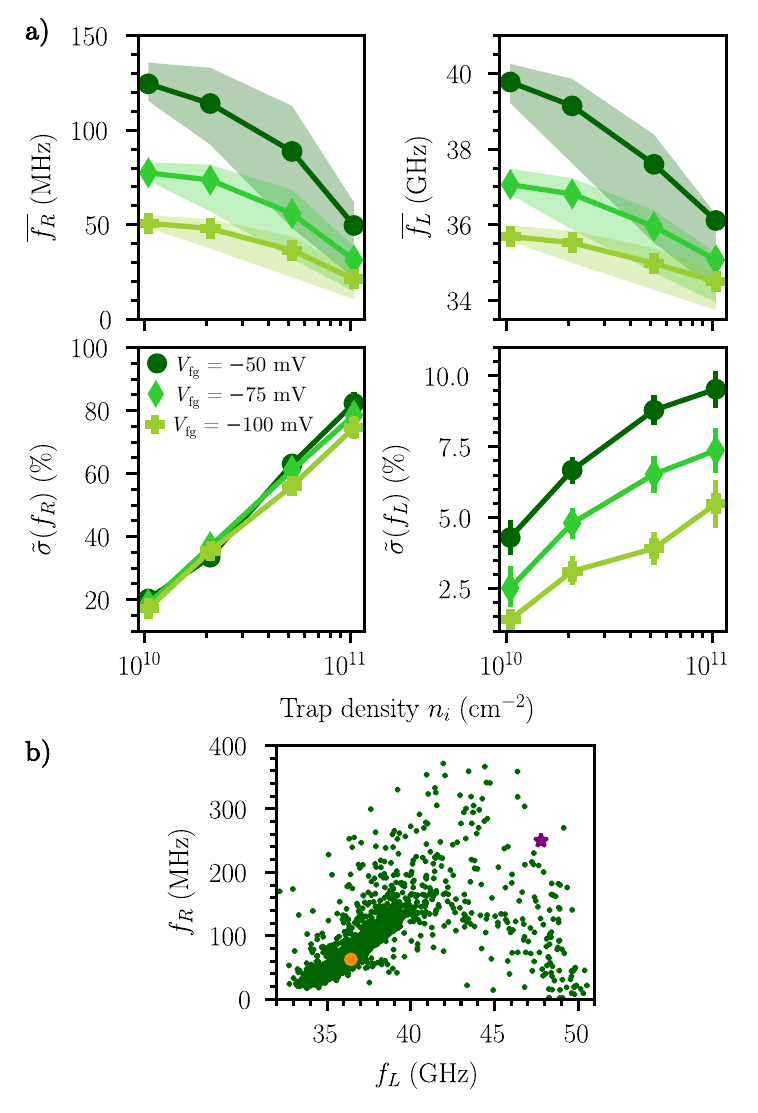}
\caption{(a) Average $\overline{f_R}$ and RSD $\tilde{\sigma}(f_R)$ of the Rabi frequency of hole qubits as a function of the density of charge traps $n_i$ at the Si/SiO$_2$ interface, for different $V_{\rm fg}$; average $\overline{f_L}$ and RSD $\tilde{\sigma}(f_L)$ of the Larmor frequency of the same hole qubits. The error bars are the 95\% confidence intervals. The first and third quartiles of the distribution of devices are also displayed as a shaded area for each $V_{\rm fg}$; 25\% of the devices lie below, 25\% above, and 50\% within this shaded area. (b) Distribution of the hole devices in the $(f_L, f_R)$ plane at $n_i=5\times 10^{10}$ cm$^{-2}$ and $V_{\rm fg}=-50$ mV. Each green point is a particular realization of the charge disorder. The orange point is the pristine device (recomputed with a homogeneous density of charges at the Si/SiO$_2$ interface), and the purple star the disordered device of Fig. \ref{fig:CTholes2}b.}
\label{fig:CTholes1}
\end{figure}

The average Rabi frequency $\overline{f_R}$, the average Larmor frequency $\overline{f_L}$, and the RSDs $\tilde{\sigma}(f_R)$ and $\tilde{\sigma}(f_L)$ of the hole qubits are plotted as a function of the density of charge traps $n_i$ in Fig. \ref{fig:CTholes1}a. The distribution of the Larmor and Rabi frequencies of the qubits is plotted in Fig. \ref{fig:CTholes1}b for $n_i=5\times 10^{10}$ cm$^{-2}$ and $V_{\rm fg}=-50$ mV. Each point represents a particular realization of 5 unit charges in the whole device of Fig. \ref{fig:device}a. The reference data $f_L^{0\prime}$ and $f_R^{0\prime}$ for the pristine device have been recomputed with a homogeneous density of charges $\sigma=n_ie$ at the Si/SiO$_2$ interface (in order to capture the average electrostatic effect of the defects), and do therefore differ from those for interface roughness.

\begin{figure}
\centering
\includegraphics[width=.95\columnwidth]{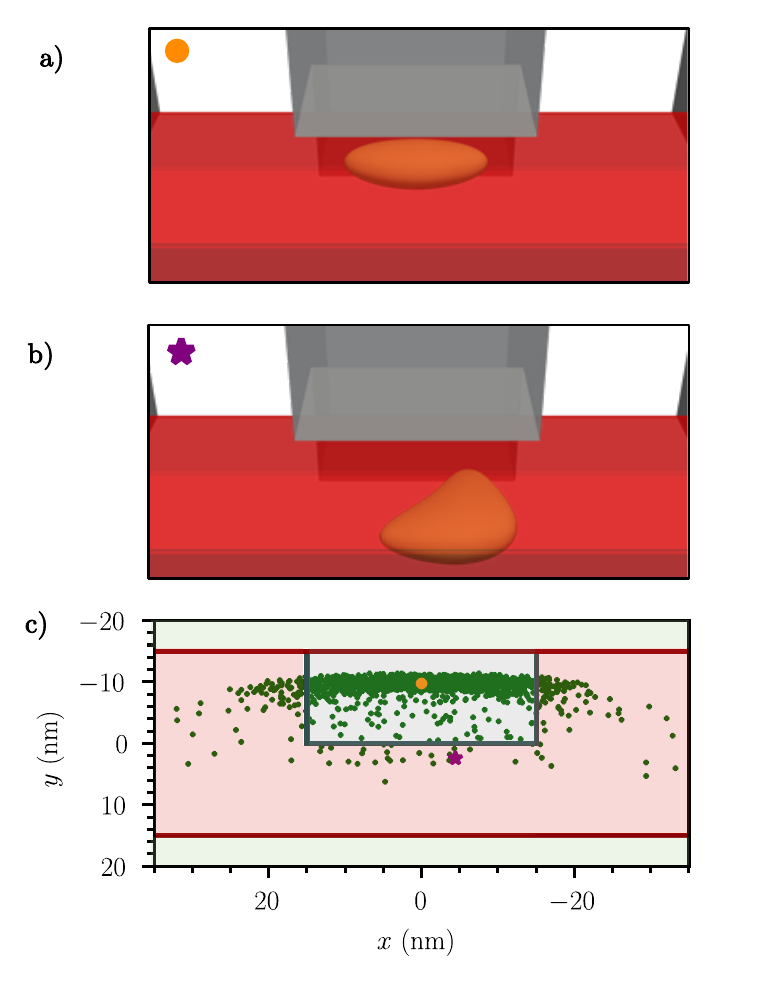}
\caption{Iso-density surface of the squared ground-state wave function of (a) the pristine hole device (orange point of  \YMN{Figs. \ref{fig:CTholes1}b and \ref{fig:CTcorrelations}}) and (b) a disordered device (purple star of \YMN{Figs. \ref{fig:CTholes1}b and \ref{fig:CTcorrelations}}) at $n_i=5\times 10^{10}$ cm$^{-2}$ and $V_{\rm fg}=-50$ mV. (c) Distribution of the average position of the ground state ($\langle x\rangle$, $\langle y\rangle$) for different realizations of the charge disorder at $n_i=5\times 10^{10}$ cm$^{-2}$ and $V_{\rm fg}=-50$ mV. The orange point and purple star are the devices of panels (a) and (b).}
\label{fig:CTholes2}
\end{figure}

The data are much scattered, the Rabi frequency spanning, in particular, more than one order of magnitude. The distribution of Rabi frequencies is non-normal at large $\tilde{\sigma}(f_R)$, so that we have added the first and third quartiles of this distribution on Fig. \ref{fig:CTholes1}a as an extra measure of the variability (25\% of the devices lie below, 25\% above, and 50\% within the shaded areas). There are also a few strong outliers with Rabi frequencies $>500$ MHz that were removed from the statistics.\footnote{We have actually removed from the statistics $\approx 3\%$ strong outliers with Rabi frequencies greater than $\hat{f}_R(0.75)+4{\rm IQR}(f_R)$. They usually result from the resonance between two coupled dots and are extremely sensitive to the bias point.} The envelope functions are, indeed, strongly shifted and distorted by the disorder, especially along the channel (the weakest confinement axis), as illustrated in Fig. \ref{fig:CTholes2}c. These real-space distortions give rise to large fluctuations of $f_L$ and $f_R$ owing to the strong spin-orbit coupling in the valence band. The displacements of the dots along the channel will also complicate the management of the exchange interactions between neighboring dots. 

\begin{figure}
\centering
\includegraphics[width=.95\columnwidth]{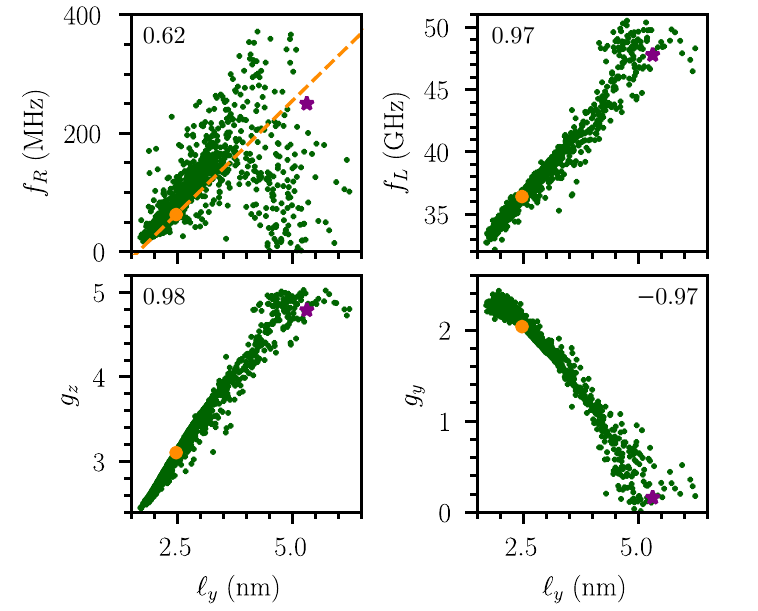}
\caption{Correlations between the Rabi frequency $f_R$, the Larmor frequency $f_L$, the gyromagnetic factors $g_y$ and $g_z$ of the hole qubits and the extension $\ell_y$ of the dots along $y$. Each point is a particular realization of charge disorder at $n_i=5\times 10^{10}$ cm$^{-2}$ and $V_{\rm fg}=-50$ mV. The orange point is the pristine device, \YMN{and the purple star the disordered device of Fig. \ref{fig:CTholes2}b. The correlation coefficient $\rho$ [Eq. (\ref{eq:corr})] between the abscissas and ordinates is given in each panel.} The dashed orange line $\delta f_R/f_R^{0\prime}=3\delta\ell_y/\ell_y^{0\prime}$ is a guide to the eye.} 
\label{fig:CTcorrelations}
\end{figure}

Remarkably, and in contrast with interface roughness, the charge traps directly modulate the lateral (rather than vertical) confinement, because the hole gets excluded from the whole thickness of the film in the vicinity of a defect ($\sigma(\ell_z)=0.01$ \AA\  at $n_i=5\times 10^{10}$ cm$^{-2}$ and $V_{\rm fg}=-50$ meV, while $\sigma(\ell_z)=0.6$ \AA\  on Fig. \ref{fig:SRcorrelations}b). As a consequence, $y$ and $z$ remain good principal axes of the $g$-tensor, so that $f_L$ shows significant correlations with $\ell_y$ despite partial cancellations between the variations of $g_y$ and $g_z$ (Fig. \ref{fig:CTcorrelations}). The deviations of the Larmor and Rabi frequencies are, therefore, both primarily dependent on $\ell_y$ and are broadly correlated (Fig. \ref{fig:CTholes1}). In general, the charge traps, which repel the holes, squeeze the dots and hinder their motion, so that the average Rabi frequency decreases with increasing $n_i$. As a matter of fact, $\overline{\ell_y}=2.66$ nm on Fig. \ref{fig:CTcorrelations}, which is smaller than $\ell_y^0=3.28$ nm in the pristine, traps-free device (table \ref{tab:fLR0H}), \YMN{but larger than $\ell_y^{0\prime}=2.47$ nm in the reference device with a homogeneous distribution of charges at the Si/SiO$_2$ interface (orange point in Figs. \ref{fig:CTholes1}--\ref{fig:CTcorrelations}). In the same way, the average $\overline{f_R}=88.8$ MHz is larger than the Rabi frequency $f_R^{0\prime}=62.8$ MHz in that reference device, which outlines the non-linear response of the spins to nearby traps (they are indeed expected equal in a first-order perturbation theory such as Appendix \ref{app:Model}).} The highest Rabi frequencies typically result from the the formation of strongly coupled multiple dots below the gate, allowing for large charge oscillations under electrical driving. 

$\tilde{\sigma}(f_R)$ decreases with decreasing $n_i$, but remains significant down to $n_i=10^{10}$ cm$^{-2}$ (only one charge trap in the whole device of Fig. \ref{fig:device}a). Indeed, a single stray defect can have a sizable effect of the nearby qubit(s), as shown in Appendix \ref{app:Additions}. As discussed in Appendix \ref{app:Model}, the scattering strength of the charged traps is expected to scale as $\sqrt{n_i}$ within first-order perturbation theory; $\tilde{\sigma}(f_R)$ actually increases faster than $\sqrt{n_i}$, but slower than $n_i$. This is presumably due to the facts that $\overline{f_R}$ also decreases with increasing $n_i$, and that $\sigma(f_R)$ is dominated by the $P_b$ defects very near or in the dot, to which the response is non linear. $\tilde{\sigma}(f_L)$ tends to saturate at large $n_i$ due to a complex interplay between the variations of the $g$-factors $g_y$ and $g_z$ of the strongly distorted dots.

\subsection{Electrons}

\begin{figure}
\centering
\includegraphics[width=.95\columnwidth]{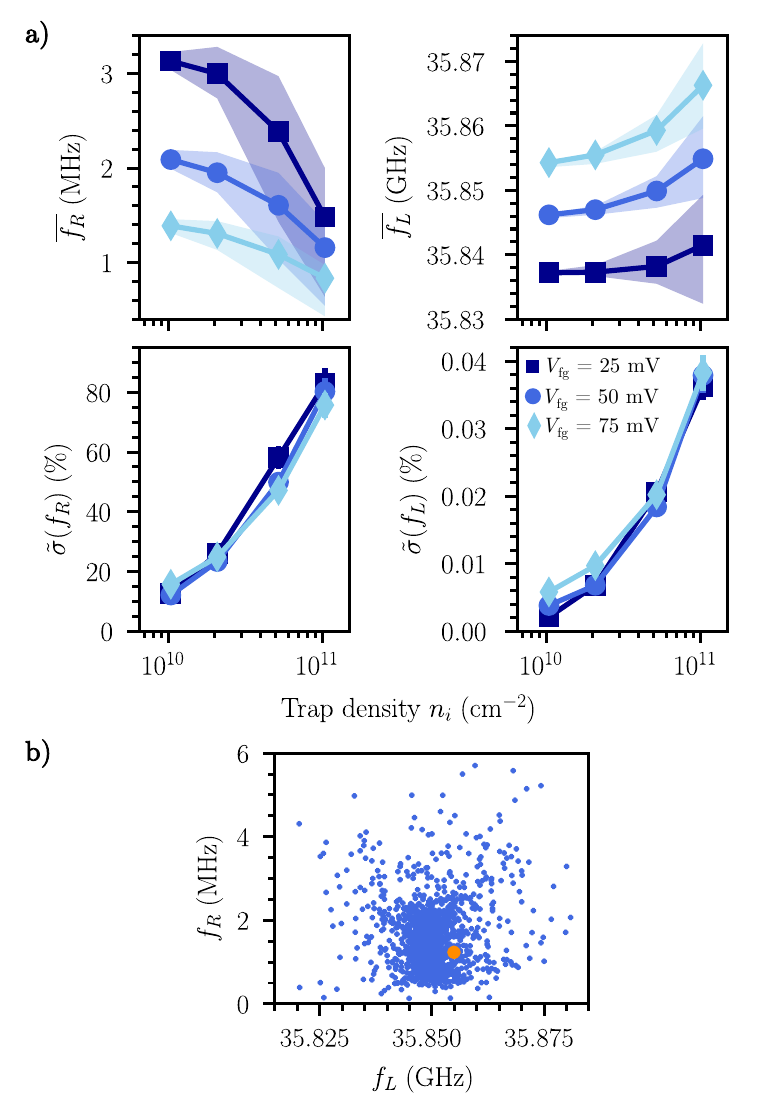}
\caption{(a) Average $\overline{f_R}$ and RSD $\tilde{\sigma}(f_R)$ of the Rabi frequency of electron qubits as a function of the density of charge traps $n_i$ at the Si/SiO$_2$ interface, for different $V_{\rm fg}$; average $\overline{f_L}$ and RSD $\tilde{\sigma}(f_L)$ of the Larmor frequency of the same electron qubits. The error bars are the 95\% confidence intervals. The first and third quartiles of the distribution of devices are also displayed as a shaded area; 25\% of the devices lie below, 25\% above, and 50\% within this shaded area. (b) Distribution of the electron devices in the $(f_L, f_R)$ plane at $n_i=5\times 10^{10}$ cm$^{-2}$ and $V_{\rm fg}=50$ mV. Each blue point is a particular realization of the charge disorder. The orange point is the pristine device.}
\label{fig:CTelectrons}
\end{figure}

The data for electrons are shown in Fig. \ref{fig:CTelectrons}. \YMN{We remind that $P_b$ defects are amphoteric, and therefore repel electrons in $n$-type devices (as they do repel holes in $p$-type devices) once charged}. The variability of the Larmor frequency is much smaller for electrons than for holes for the same reasons as for interface roughness \YMN{(weak dependence of $B_y$ on vertical confinement). The average Larmor frequency $\overline{f_L}$ slightly increases with increasing $n_i$ as the electron tends to move upward in the channel (the charged defects being less screened by the gate on the bottom than on the top interface)}. The variability of the Rabi frequency is, nonetheless, as large as for holes, especially at the highest trap densities. The potential of the charge traps is, indeed, the same for electrons and holes, in contrast with the effective potential for interface roughness, which depends on the confinement mass of the carriers [Eq. (\ref{eq:WSR})]. \YMN{Also, this potential is expected to have similar effects on the in-plane motion of electrons and holes, since the in-plane mass $m_\parallel^*\simeq 0.2\,m_0$ of both carriers is comparable.\footnote{\YMN{In fact, the charge traps have a stronger effect on the in-plane motion of holes (larger $\tilde{\sigma}(\ell_y)$) owing to the multi-bands character of the Hamiltonian. Yet the scaling of the Rabi frequency is softer for holes ($\delta f_R/f_R^{0\prime}\approx 3\delta\ell_y/\ell_y^{0\prime}$) than for electrons ($\delta f_R/f_R^{0\prime}\approx 4\delta\ell_y/\ell_y^{0\prime}$). Therefore, the net impact of charged traps is about the same for electrons and holes.}}} The Rabi frequency of electrons remains largely correlated to $\ell_y$, except for some specific trap configurations where the dot is strongly displaced or squeezed, or, on the opposite, where coupled dots form under the gate as a result of the disorder.

The charge disorder variability of the Larmor and Rabi frequencies of both electron and hole qubits shows only a weak dependence on the channel thickness $H$.

\section{Discussions}
\label{sec:Discussion}

We now outline the implications of the above results and possible strategies to mitigate disorder. We discuss, in particular, the relations between the variability of the Rabi frequencies and of the qubit lifetimes ($T_1$, $T_2$), and the consequences for multi-qubit systems.

\subsection{Sensitivity to disorder}
\label{sec:Sensitivity}

Spin-orbit qubits are sensitive to disorder. This is the price to pay for the possibility to manipulate the spins electrically. Unless material and device engineering can mitigate the disorder, each individual qubit needs to be characterized separately in order to account for its own ``personality''. On the other hand, the same spin-electric coupling provides opportunities to tune the qubits and partly correct for device-to-device deviations. This, of course, requires that the spread of characteristics is manageable, and that the qubits remain functional and can be coupled together.

In this respect, charge traps are much more critical than interface roughness. This is reminiscent of the behavior of classical field effect transistors. Indeed, a spin qubit at a Si/SiO$_2$ interface is in essence a field-effect transistor operating in the low electric field/low carrier density range; the carrier mobility is well known to be primarily limited by the (unscreened) Coulomb disorder in this regime.\cite{Esseni03,Niquet14,Nguyen14} Disorder is an even stronger concern for qubits as the relevant energy scales are typically in the sub-meV range (as compared to the $kT$ or source-drain voltage range for classical transport) while fluctuations are in the few meV range.

\YMN{We would like to draw attention to the fact that electron spin qubits are subject to additional sources of variability not accounted for in this work. First, spin-valley-orbit coupling (SVOC) is neglected in the present effective mass approximation. Interface roughness is known to be responsible for a significant variability of the valley splitting and spin-valley mixing, as discussed for example in Refs. \onlinecite{Culcer10,Bourdet18}. This is not expected to have a strong impact on the Rabi frequencies (the dipole matrix elements between valley states being small along the main direction $y$ of the EDSR motion), unless the valley and Zeeman splittings are close enough to allow for intrinsic SVOC-driven Rabi oscillations.\cite{Corna18,Bourdet18} However, SVOC may slightly lower the $g$-factor of electrons (by up to a few hundredths);\cite{Veldhorst15,Ruskov18,Ferdous18,Tanttu19} device to device fluctuations of the $g$-factor $\delta g=0.005$ would give rise to variations of the Larmor frequency $\delta f_L=70$ MHz at a net field $B=1$ T. Such fluctuations are on the scale or even larger than those reported in Figs. \ref{fig:SRElectrons} and \ref{fig:CTelectrons}. Achieving robust and controllable valley splitting is actually a key to the realization of well defined two-level systems for spin manipulation and readout. Also, the electron spin qubits may be sensitive to local inhomogeneities (roughness, variations of the magnetic polarization...) and global misalignment (misplacement/misorientation) of the micro-magnets.\cite{Yoneda15,Simion20} These disorders, which are specific of extrinsic spin-orbit coupling, are addressed in Appendix \ref{app:Varmicro}.}

As pointed out in section \ref{sec:Methodology}, we may attempt to drive the hole devices with the two side gates on the left and right of the dot (the present micro-magnets layout is not suitable for that purpose in the electron qubits). The dot then essentially moves as a whole along the channel axis $x$. As discussed in Ref. \onlinecite{Michal21}, the Rabi frequency in this ``iso-Zeeman'' mode is proportional to $\ell_x^4$. The situation is not, therefore, expected to be more favourable: the channel axis being typically the direction of weakest confinement, $\ell_x$ shows strong device-to-device variations (as evidenced in Figs. \ref{fig:SRcorrelations} and \ref{fig:CTholes2}). This conclusion is supported by numerical simulations showing comparable or even larger Rabi frequency variability in this driving mode (see Appendix \ref{app:Additions}). As a matter of fact, these simulations also show the emergence of a significant $g$-TMR contribution on top of the original iso-Zeeman one, the disorder giving rise to sizable modulations of the gyromagnetic factors under driving.

Finally, we would like to emphasize that the present data were collected at fixed gate voltage. We may, alternatively, collect data at fixed chemical potential $\mu$ in the dot (fixed ground-state energy), with the prospect of reducing the variability (because $\mu$ shall also correlate with the extension of the ground-state wave function). \YMN{This is moreover closer to the experimental situation when the dots remain connected to reservoirs of particles while being operated.} The computational procedure is, however, more complex and time-consuming, as the bias must be corrected for each disordered device in order to achieve the target chemical potential. Yet our attempts for interface roughness and charge traps did not show any clear improvement of the variability. As a matter of fact, even in a separable 3D model, working at fixed chemical potential does not ensure that the $x$, $y$, $z$ motion energies are all preserved (only their sum is). As suggested above, the variability can indeed be partly compensated by bias adjustments, however aimed at correcting, e.g., the extension along $y$ or the position along $z$, and not the total energy of the dot. Such bias corrections may nonetheless be limited by the stability diagram of the devices. \YMN{To be more specific, the Larmor frequency of 46\% of the rough hole qubits, and of 55\% of the hole quits with charged traps can be matched to the pristine device with bias corrections $|\Delta V_{\rm fg}|\le 20$ mV ($V_{\rm fg}=-50$ mV; $\Delta=0.4$ nm, $L_c=10$ nm; $n_i=5\times 10^{10}$ cm$^{-2}$). This $|\Delta V_{\rm fg}|=20$ mV is however already about twice the typical charging energy and therefore calls for tight control over the barriers to prevent inter-dot tunneling. The situation is not much better for electrons despite the smaller variability of the Larmor frequency, as the Stark effect (electrical tunability of $f_L$) is much weaker than for holes.}

\subsection{Relations with qubit lifetimes}

\begin{figure}
\centering
\includegraphics[width=.95\columnwidth]{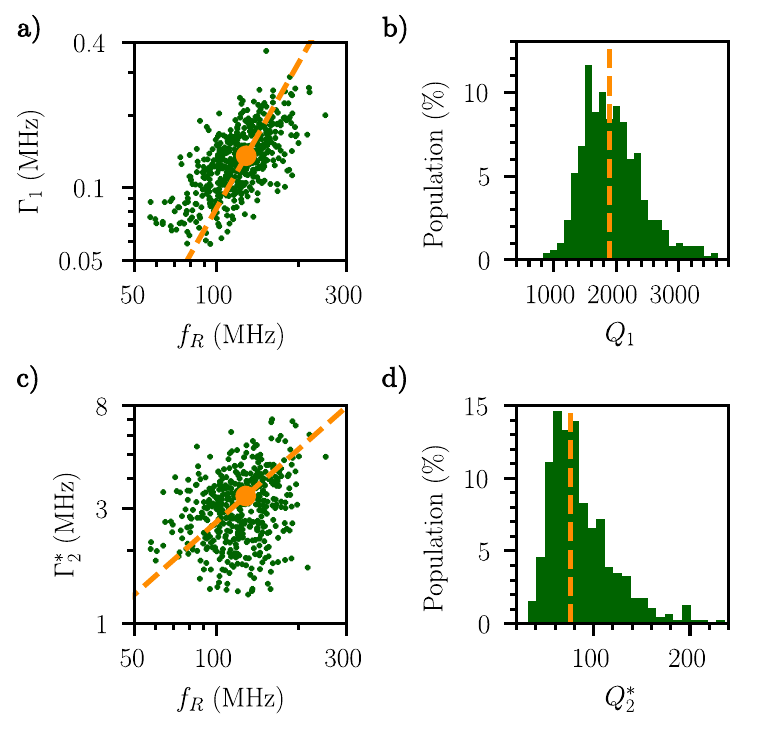}
\caption{(a) Distribution of the rough hole devices in the $(f_R, \Gamma_1)$ plane ($\Delta=0.4$ nm, $L_c=10$ nm, $V_{\rm fg}=-50$ mV and $B_{\rm ext}=1$ T). Each green point is a particular realization of the interface roughness disorder. The orange point is the pristine device. The dashed orange line, $\Gamma_1\propto(f_R/f_R^0)^2$, is provided as a guide to the eye. (b) Histogram of the quality factors $Q_1=f_R\times T_1$ for the same set of devices. The orange dashed line is the quality factor of the pristine device. (c) Distribution of the rough hole devices in the $(f_R, \Gamma_2^*)$ plane (same conditions as before). The dashed orange line, $\Gamma_2^*\propto f_R/f_R^0$, is provided as a guide to the eye. (d) Histogram of the quality factors $Q_2^*=f_R\times T_2^*$ for the same set of devices.}
\label{fig:T1}
\end{figure}

The disorder will also give rise to variability in the qubit relaxation time $T_1$ and dephasing time $T_2^*$. This can have detrimental consequences on the performances of a multi-qubits processor.

First, we expect as a general trend that the devices with the largest Rabi frequencies also show the shortest relaxation times $T_1$. Indeed, $\Gamma_1=1/T_1\propto\sum f_L^n|M_1|^2$ in the Fermi-Golden rule/Bloch-Redfield approximation, where the exponent $n$ depends on the relaxation mechanism ($n=1$ for Johnson-Nyquist noise, $n=3$ to $5$ for phonons\cite{Paladino14,Tahan14,Hu14,Li20}), and $M_1\equiv\bra{\Uparrow}{\cal M}_1\ket{\Downarrow}$ is the matrix element of an operator ${\cal M}_1$ that describes some spin-electric coupling. $M_1$ can, therefore, be expected to scale with $f_R$ (which is also such a transverse spin-electric coupling matrix element).

The pure dephasing rate is, likewise, $\Gamma_2^*=1/T_2^*\propto \sum |M_2|^m$, where  $M_2\equiv\bra{\Uparrow}{\cal M}_2\ket{\Uparrow}-\bra{\Downarrow}{\cal M}_2\ket{\Downarrow}$ for some coupling operator ${\cal M}_2$, $m=2$ for regular noise (Bloch-Redfield approximation) and $m=1$ for quasi-static $1/f$ noise.\cite{Paladino14} Although the relations between the longitudinal matrix elements involved in $\Gamma_2^*$ and the transverse matrix elements involved in $f_R$ and $\Gamma_1$ is far from obvious, we can still expect that devices with stronger spin-electric coupling show, on average, larger $f_R$, $\Gamma_1$, and $\Gamma_2^*$, unless some sweet spot has been found. Some $P_b$ defects themselves may be a source of $1/f$ noise \YMN{if their charge trapping/detrapping or structural rearrangement times are shorter than the duration of an experiment.\cite{deSousa76}} This problem goes, however, beyond the scope of the present work.

We illustrate these trends on the relaxation time $T_1$ and on the pure dephasing time $T_2^*$ of holes. The phonon-limited $\Gamma_1$ is calculated along the lines of Ref. \onlinecite{Li20}. As for $\Gamma_2^*$, we assume a quasi-static $1/f$ noise with finite bandwidth whose action on the qubit can be modeled as an effective fluctuation $\delta V_{\rm fg}(t)$ of the gate voltage with rms amplitude $\delta V_{\rm fg,rms}$. Then,\cite{Dial13}
\begin{equation}
\Gamma_2^*=\frac{1}{\sqrt{2}\hbar}e|\bra{\Uparrow}D\ket{\Uparrow}-\bra{\Downarrow}D\ket{\Downarrow}|\delta V_{\rm fg,rms}\,,
\end{equation}
where $D(\vec{r})$ is the derivative of the total potential $V(\vec{r})$ in the device with respect to the gate voltage $V_{\rm fg}$. We set $\delta V_{\rm fg,rms}=5$ $\mu$V as an illustration. Note that for such a $1/f$ noise the coherence decays as $\exp[-(t/T_2^*)^2]$ instead of $\exp[-t/T_2^*]$.

The distribution of Rabi frequencies $f_R$ and relaxation rates $\Gamma_1$ is plotted in Fig. \ref{fig:T1}a for interface roughness disorder ($\Delta=0.4$ nm, $L_c=10$ nm). The corresponding data for $\Gamma_2^*$ are plotted in Fig. \ref{fig:T1}c. They were computed at $B_{\rm ext}=1$ T along $\vec{y}+\vec{z}$ ($\Gamma_1$ scales\cite{Li20} as $B_{\rm ext}^5$ and $\Gamma_2^*$ as $B_{\rm ext}$). As hinted above, the larger the Rabi frequency, the larger $\Gamma_1$ (and, to a much lesser extent, $\Gamma_2^*$) tends to be. There is, nonetheless, a significant spread of the single qubit quality factors $Q_1=2f_R\times T_1$ and $Q_2^*=2f_R\times T_2^*$ (number of $\pi$ rotations that can be achieved within $T_1$ or $T_2^*$), as shown in Fig. \ref{fig:T1}b, d.

In an ensemble of qubits, the relevant figure of merit for relaxation is however ${\hat Q}_1=2\min(f_R)\times\min(T_1)$. It is limited by the slowest and by the shortest-lived qubits, which are in principle different. Assuming $\Gamma_1\propto (f_R/f_R^0)^\alpha$, we can give an estimate for ${\hat Q}_1$:
\begin{equation}
{\hat Q}_1\approx Q_1^0\frac{\min(f_R)/f_R^0}{\left[\max(f_R)/f_R^0\right]^\alpha}\,,
\end{equation}
where $Q_1^0$ is the quality factor of the pristine device. A similar expression can be obtained for $\hat{Q}_2^*=2\min(f_R)\times\min(T_2^*)$, with a possibly different $\alpha$. This highlights how detrimental the variability can be for the operation of an ensemble of qubits. In the case of Fig. \ref{fig:T1}, ${\hat Q}_1=444$ on the 90\% best qubits (450 best individual $Q_1$ out of 500 devices), and ${\hat Q}_2^*=16.7$, much lower than $Q_1^0=1895$ and $Q_2^{*0}=76.0$. The operation of slow qubits can be sped up by increasing the driving RF power (${\hat Q}\to Q^0/[\max(f_R)/f_R^0]^\alpha$ at same power-corrected Rabi frequencies), yet at the expense of a more complex RF management on the chip, and at the risk of heating up the qubits.

\subsection{Mitigation of the disorder}

The qubits may be made more resilient to variability through material and/or device engineering. In particular, the previous sections highlight how critical is the quality of materials and interfaces for the control and reproducibility of spin qubits. Improving the smoothness and passivation of the Si/SiO$_2$ interface will definitely reduce variability; a RSD $\tilde{\sigma}(f_R)<10\%$ however calls for very clean and stable interfaces with charged defect densities $n_i<10^{10}$ cm$^{-2}$ that are at the state-of-the-art. 

The interface roughness variability can be largely alleviated by a proper optimization of the channel/film thickness and vertical electric field in order to reach the best balance between single qubit performances and sensitivity to disorder (see Fig. \ref{fig:Thickness}). This optimum is presumably dependent on the device layout and mechanisms used to drive the spin.

\begin{figure}
\centering
\includegraphics[width=.95\columnwidth]{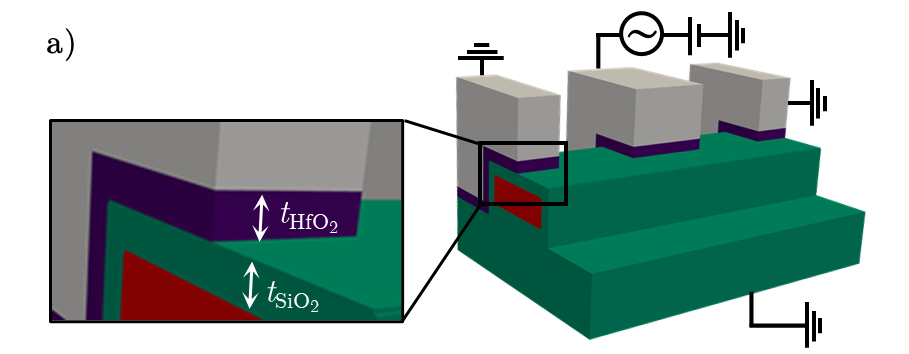}
\includegraphics[width=.95\columnwidth]{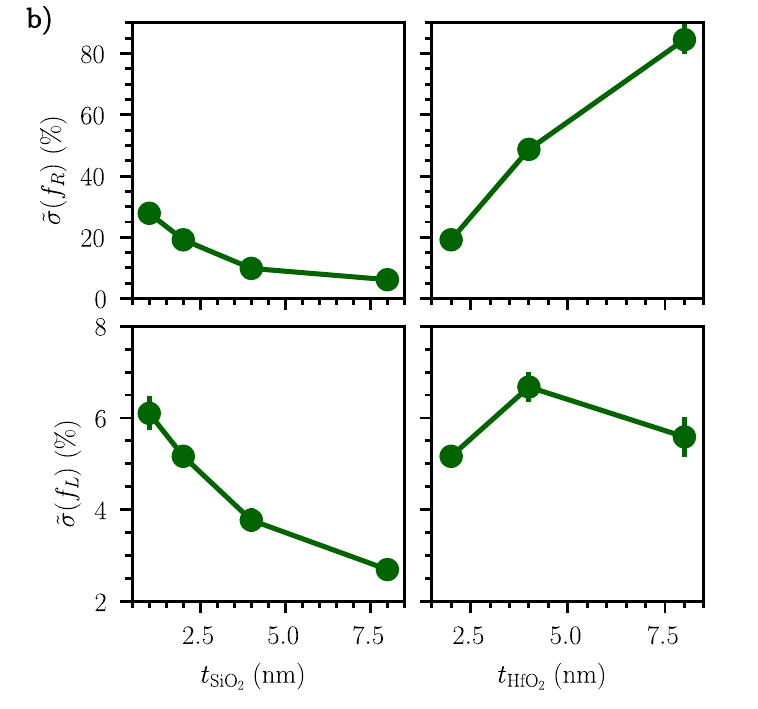}
\caption{(a) Hole device with a SiO$_2$/HfO$_2$ gate stack. The thickness of the SiO$_2$ layer (green) is $t_{{\rm SiO}_2}$, and the thickness of the HfO$_2$ layer (purple) is $t_{{\rm HfO}_2}$. The HfO$_2$ layer extends only under the gates. (b) RSDs $\tilde{\sigma}(f_L)$ and $\tilde{\sigma}(f_R)$ of the Larmor and Rabi frequencies plotted as a function of $t_{{\rm SiO}_2}$ (at $t_{{\rm HfO}_2}=2$ nm), and as a function of $t_{{\rm HfO}_2}$ (at $t_{{\rm SiO}_2}=2$ nm). The density of trapped charges at the SiO$_2$/HfO$_2$ interface is $n_i=5\times10^{11}$ cm$^{-2}$, and the bias voltage is chosen so that the ground-state energy of the pristine device remains the same as in the original qubit of Fig. \ref{fig:device} at $V_{\rm fg}=-50$ mV whatever $t_{{\rm SiO}_2}$ and $t_{{\rm HfO}_2}$. The error bars are the 95\% confidence intervals.}
\label{fig:CTSiO2}
\end{figure}

Nevertheless, one of the most reliable solutions to the variability problem is to switch from a crystalline/amorphous interface such as Si/SiO$_2$ to an epitaxial interface such as Si/SiGe (electron qubits) or Ge/SiGe (hole qubits). Indeed, epitaxial interfaces show, in principle, low roughness and very small density of traps. Strains must be carefully addressed in order to avoid threading dislocations in the active areas, but this is now fluently managed in group IV materials. The four hole qubits device of Ref. \onlinecite{Hendrickx21} was actually realized in a Ge/SiGe quantum well controlled by accumulation gates.

The charged defects are then deported at the surface of the heterostructures on which the metal gates are deposited. This surface is usually no more than 50 nm away from the well (in order to keep a tight enough electrostatic control on the qubits). We emphasize, though, that the impact of charge traps decreases significantly once they are taken away from the quantum dots. As an illustration, we have considered SOI hole devices with a SiO$_2$/HfO$_2$ gate stack (Fig. \ref{fig:CTSiO2}a). The channel is therefore now separated from the gate by a layer of SiO$_2$ with thickness $t_{{\rm SiO}_2}$, and by a layer of HfO$_2$ with thickness $t_{\rm HfO_2}$ ($\kappa_{{\rm HfO}_2}=20$). This HfO$_2$ layer only extends below the gates and not under the spacers. We then introduce the charged defects at the SiO$_2$/HfO$_2$ instead of the Si/SiO$_2$ interface, with density $n_i=5\times10^{11}$ cm$^{-2}$ only chosen\footnote{We emphasize that the SiO$_2$/HfO$_2$ interface is known to be a strong source of Coulomb scattering in classical CMOS devices,\cite{Zeng17} with apparent charge densities $n_i$ reaching $10^{13}$ cm$^{-2}$. This test system is introduced to illustrate trends in a device similar to section \ref{sec:Chargetraps}, and is not meant to give a realistic account of disorder at the SiO$_2$/HfO$_2$ interfaces, which shall preferably be avoided in qubit devices.} for illustrative purposes.\cite{Zeng17} The RSDs $\tilde{\sigma}(f_L)$ and $\tilde{\sigma}(f_R)$ of the Larmor and Rabi frequencies are plotted as a function of $t_{{\rm SiO}_2}$ and $t_{{\rm HfO}_2}$ in Fig. \ref{fig:CTSiO2}b. The variability decreases when the SiO$_2$ is made thicker and the traps are moved away from the channel. Remarkably, the variability increases rapidly with the thickness of the HfO$_2$ layer because the screening of the charge traps by the metal gate is softened.\cite{Zeng17} In general, surrounding the qubits by materials with higher dielectric constant (SiGe $vs$ SiO$_2$), and by a dense set of gates will reduce the impact of charged defects on variability (and possibly of charge noise on qubit lifetimes).\footnote{This ``electrostatic'' argument assumes that these materials are not themselves sources of additional static or dynamic noise.} Working in the many electrons/holes regime may also enhance screening, but usually makes the dots larger and more responsive to disorder. The optimal number of particles in the dots (as far as variability is concerned) remains, therefore, an open question.\cite{Leon20}

Finally, the model of Appendix \ref{app:Model} shows that the variability increases with the in-plane mass $m_\parallel^*$ (at given dot sizes $\ell_x^0$ and $\ell_y^0$). Heavier particles indeed localize more efficiently in the disorder. It is, therefore, {\it a priori} advantageous to switch from silicon to lighter mass materials such as germanium for holes (for interface roughness, the variability is $\propto m_\parallel^*/m_\perp^*$ when $H\ll\ell_{\mathcal{E}_z}$ so that Ge is advantageous over Si even in this regime). However, the dots are usually made larger in light mass materials (the dot sizes $\ell_x^0$ and $\ell_y^0$ scale as $(m_\parallel^*)^{-1/4}$ in a given parabolic potential for example, and as $(m_\parallel^*)^{-1/2}$ at given confinement energy), hence can be more polarizable and sensitive to disorder. Eq. (\ref{eq:harmonic}) actually suggests that the variability may not necessarily improve when decreasing the mass at given confinement energy, especially in the presence of long-wavelength disorders such as charge traps.

\section{Conclusions}

We have investigated the variability of single qubit properties (Larmor and Rabi frequencies) due to disorder at the Si/SiO$_2$ interface (roughness and charge traps) in MOS-like devices. We have, in particular, compared hole qubits subject to intrinsic spin-orbit coupling and electron qubits subject to a synthetic spin-orbit coupling created by micro-magnets. The Larmor frequencies of electrons are more robust to disorder than the Larmor frequencies of holes, which are rather sensitive to changes in the potential landscape. The Rabi frequencies show anyway much larger variability than the Larmor frequencies for both kinds of carriers. The deviations of the Rabi frequencies can be traced back to the modulations of the size of the dots by the disorder. In thin $(001)$ films, holes are more sensitive to interface roughness than electrons because the confinement mass of the heavy-holes is smaller than the confinement mass of the electrons in the Z-valleys (hence the effects of fluctuations of the film thickness are larger). The main source of variability is, however, charge traps at the Si/SiO$_2$ interface, which can spread both electron and hole Rabi frequencies over one order of magnitude. The dots can be significantly distorted and displaced by charge disorder, which does not only scatter one qubit properties, but may also complicate the management of exchange interactions between the dots. The disorder also scatters the relaxation and dephasing times, which systematically degrades the figures of merit of an ensemble of qubits (as the lifetimes of such an ensemble are limited by the poorest qubits). Low variability $\tilde{\sigma}(f_R)<10\%$ calls for smooth and clean interfaces with charge traps densities $n_i<10^{10}$ cm$^{-2}$. This is presumably more easily achieved with epitaxial heterostructures such as Si/SiGe or Ge/SiGe, where the residual (surface) charge traps can be deported tens of nanometers away from the active layer. The impact of charge traps indeed decreases very fast once they are moved away from the qubits, especially when the latter are embedded in materials with high dielectric constants and are controlled by dense sets of gates that screen the Coulomb disorder.

\section*{Acknowledgements}

We thank Michele Filippone for fruitful comments and suggestions. This project was supported by the European Union's Horizon 2020 research and innovation programme under grant agreement 951852 (project QLSI), and by the French national research agency (project MAQSi).

\appendix

\section{Vector potential created by the micro-magnets}
\label{app:micromagnets}

The magnetic field created by micro-magnets has been discussed, for example, in Refs. \onlinecite{Neumann15}, \onlinecite{Yang90}, \onlinecite{Goldman00} and \onlinecite{Yoneda15}. Here we deal with the calculation of the corresponding vector potential, which is the natural input for quantum mechanics codes (see appendix \ref{app:Rabi}).

We consider a micro-magnet layout that can be viewed as an arrangement of homogeneous bar magnets. The total vector potential and magnetic field are hence the sum of those of each individual bar magnet. Let ${\cal B}$ be a bar magnet with magnetic moment density $\vec{M}=M\vec{w}$ and sides $L_u$, $L_v$ and $L_w$ along three orthonormal axes $\vec{u}$, $\vec{v}$, and $\vec{w}$. The vector potential $\vec{A}$ created at point $\vec{r}$ by this magnet is:
\begin{equation}
\vec{A}(\vec{r})=\frac{\mu_0}{4\pi}\vec{M}\times\int_{\cal B} d^3\vec{r}'\,\frac{\vec{r}-\vec{r}'}{|\vec{r}-\vec{r}'|^3}\,.
\end{equation}
Setting $\vec{r}=(u,v,w)$ in the $\{\vec{u},\vec{v},\vec{w}\}$ axis set with the origin at the center of the magnet, explicit integration yields:
\begin{align}
A_\alpha&(u,v,w)=\frac{J}{4\pi}\sum_{i,j,k=0}^{1} (-1)^{i+j+k} \nonumber \\
&F_\alpha\left(u+(-1)^i\frac{L_u}{2}, v+(-1)^j\frac{L_v}{2}, w+(-1)^k\frac{L_w}{2}\right)\,,
\end{align}
where $\alpha\in\{u,v,w\}$ and $J=\mu_0M$ is the magnetic polarization. The $F$ functions are:
\begin{subequations}
\begin{align}
F_u(U,V,W)&=-V\text{atan}\left(\frac{UW}{VR}\right) \nonumber \\
&+W\text{ln}\left(R+U\right)+U\text{ln}\left(R+W\right) \\
F_v(U,V,W)&=U\text{atan}\left(\frac{VW}{UR}\right) \nonumber \\
&-W\text{ln}\left(R+V\right)-V\text{ln}\left(R+W\right) \\
F_w(U,V,W)&=0
\end{align}
\end{subequations}
with $R=\sqrt{U^2+V^2+W^2}$. The components of the magnetic field $\vec{B}=\vec{\nabla}\times\vec{A}$ then read:
\begin{align}
B_\alpha&(u,v,w)=\frac{J}{4\pi}\sum_{i,j,k=0}^{1} (-1)^{i+j+k} \nonumber \\
&G_\alpha\left(u+(-1)^i\frac{L_u}{2}, v+(-1)^j\frac{L_v}{2}, w+(-1)^k\frac{L_w}{2}\right)
\end{align}
with:
\begin{subequations}
\begin{align}
G_u(U,V,W)&=\text{ln}\left(R+V\right) \\
G_v(U,V,W)&=\text{ln}\left(R+U\right) \\
G_w(U,V,W)&=-\text{atan}\left(\frac{UV}{WR}\right)\,.
\end{align}
\end{subequations}
The above expressions for the vector potential and magnetic field are valid outside the magnet.

In the present calculations, $L_y\gg L_x\gg L_z$ and we assume that the magnetic polarization\cite{Neumann15} of Cobalt $J=1.84$ T is saturated in an external magnetic field $B_{\rm ext}=1$ T along $\vec{y}$. The total vector potential $\vec{A}=\vec{A}_1+\vec{A}_2+\vec{A}_{\rm ext}$ is the sum of the vector potentials $\vec{A}_1$ and $\vec{A}_2$ of the two micro-magnets, and of the vector potential $\vec{A}_{\rm ext}=-\vec{r}\times\vec{B}_{\rm ext}/2$ of the external magnetic field.

\section{Rabi frequency in a gradient of magnetic field}
\label{app:Rabi}

We compute the Rabi frequencies of the electrons with an anisotropic effective mass model for the $Z$ valleys. We make the substitution $\hbar\vec{k}\to{-i\vec{\nabla}}+e\vec{A}$ (with $\vec{k}$ the wave vector) and add the spin Zeeman Hamiltonian:
\begin{equation}
H_Z=\frac{1}{2}g_0\mu_B(\vec{\nabla}\times\vec{A})\cdot\vec{\sigma}\,,    
\end{equation}
where $g_0\simeq 2$, $\mu_B$ is Bohr's magneton and $\vec{\sigma}$ is the vector of Pauli matrices. We next compute the energies $E_{\Downarrow/\Uparrow}$ and the wave functions of the the ground-state $\ket{\Downarrow}$ and first excited state $\ket{\Uparrow}$ in the electrostatic potential $V(\vec{r})$ of the gates with a finite-differences method. When a time-dependent modulation $\delta V_{\rm fg}(t)=V_{\rm RF}\sin(2\pi f_L t)$ is applied to the front gate, resonant with the Zeeman splitting $E_Z=hf_L=E_\Uparrow-E_\Downarrow$, the spin rotates at Rabi frequency:\cite{Bourdet18,Venitucci18}
\begin{equation}
f_R=\frac{e}{h}V_{\rm RF}\left|\bra{\Uparrow}D\ket{\Downarrow}\right|\,, 
\label{eq:Rabidirect}
\end{equation}
where $D(\vec{r})=\partial V(\vec{r})/\partial V_{\rm fg}$ is the derivative of the electrostatic potential $V(\vec{r})$ in the device with respect to the front gate voltage $V_{\rm fg}$ (the potential created by a unit potential on that gate while all others are grounded when the electrostatics is linear, as is the case here). 

When the gradient of magnetic field is homogeneous enough, we can derive an alternative formulation for the Rabi frequency that emphasizes its dependence on the electric dipole of the dot. We hence assume that the magnetic field $\vec{B}_{\rm m}$ created by the micro-magnets can be approximated near the dot as:
\begin{equation}
\vec{B}_{\rm m}(\vec{r})\simeq\vec{B}_{\rm m}(\vec{r}_0)+G(\vec{r}-\vec{r}_0)\,,
\end{equation}
where $\vec{r}_0$ is some reference point and $G$ is the matrix of derivatives of the magnetic field components $B_{{\rm m}\alpha}$ ($\alpha\in\{x,y,z\}$):
\begin{equation}
G_{\alpha\beta}=\frac{\partial B_{{\rm m}\alpha}}{\partial r_\beta}(\vec{r}_0)\,.
\end{equation}
Note that the $G_{\alpha\beta}$'s are not independent as they must fulfill Maxwell's equations $\vec{\nabla}\cdot\vec{B}_{\rm m}=0$ and $\vec{\nabla}\times\vec{B}_{\rm m}=\vec{0}$ outside the magnets, namely:
\begin{subequations}
\begin{align}
&G_{xx}+G_{yy}+G_{zz}=0\,, \\
&G_{\alpha\beta}=G_{\beta\alpha}\ (\alpha\ne\beta)\,.
\end{align}
\end{subequations}

Neglecting the action of the vector potential on the orbital motion, the effective Hamiltonian of the spin can then be written:
\begin{align}
H&=\frac{1}{2}g_0\mu_B\left(\vec{B}_{\rm ext}+\langle\vec{B}_{\rm m}\rangle\right)\cdot\vec{\sigma} \nonumber \\
&+\frac{1}{2}g_0\mu_B\delta V_{\rm fg}(t)\left(G \frac{\partial\langle\vec{r}\rangle}{\partial V_{\rm fg}}\right)\cdot\vec{\sigma}\,.
\end{align}
Here, $\langle\vec{r}\rangle=\bra{\Uparrow}\vec{r}\ket{\Uparrow}=\bra{\Downarrow}\vec{r}\ket{\Downarrow}$ is the average position of the dot in the ground-state at zero magnetic field, $\langle\vec{B}_{\rm m}\rangle=\vec{B}_{\rm m}(\vec{r}_0)+G(\langle\vec{r}\rangle-\vec{r}_0)$ is the average micro-magnet field seen by the dot, and $\delta V_{\rm fg}(t)=V_{\rm RF}\sin(\omega t)$ is the time-dependent modulation of the gate voltage $V_{\rm fg}$ that drives the spin. When $\hbar\omega$ is resonant with the Zeemann splitting $E_Z=g_0\mu_B|\vec{B}_{\rm ext}+\langle\vec{B}_{\rm m}\rangle|$, the spin rotates at Rabi frequency:
\begin{equation}
f_R=\frac{V_{\rm RF}}{2h}g_0\mu_B\left|\vec{b}\times\left(G \frac{\partial\langle\vec{r}\rangle}{\partial V_{\rm fg}}\right)\right|\,,
\end{equation}
where $\vec{b}$ is the unit vector aligned with the total magnetic field $\vec{B}=\vec{B}_{\rm ext}+\langle\vec{B}_{\rm m}\rangle$. The derivation is similar to the $g$-matrix formula for intrinsic spin-orbit coupling.\cite{Venitucci18} 

The response $\partial\langle\vec{r}\rangle/\partial V_{\rm fg}$ can be evaluated using either perturbation theory\cite{Pioro-Ladriere08} or finite differences at two biases $V_{\rm fg}$ and $V_{\rm fg}+\delta V$:
\begin{equation}
\frac{\partial\langle\vec{r}\rangle}{\partial V_{\rm fg}}\simeq\frac{\langle\vec{r}\rangle(V_{\rm fg}+\delta V)-\langle\vec{r}\rangle(V_{\rm fg})}{\delta V}\,.
\end{equation}
Although the above estimation is not necessarily much faster than Eq. (\ref{eq:Rabidirect}) for a single orientation of the magnetic field $\vec{B}_{\rm ext}$, it clearly highlights the relation between the Rabi frequency and the electrical response of the dot. In the present devices, $\langle\vec{B}_m\rangle$ and the $G$ matrix read at the average position of the pristine dot ($V_{\rm fg}=50$ mV): 
\begin{subequations}
\begin{align}
\langle\vec{B}_m\rangle&=(0.00,0.28,-0.01)\ {\rm T}\,, \\
G&=\begin{pmatrix}
0.00 & 0.00  & 0.00 \\
0.00 & 0.12  & 2.56 \\
0.00 & 2.56 & -0.12 
\end{pmatrix}\ {\rm mT/nm}.
\end{align}
\end{subequations}
Therefore,
\begin{equation}
f_R\propto\frac{\partial B_z}{\partial y}\frac{\partial\langle y\rangle}{\partial V_{\rm fg}}\,, 
\end{equation}
because the dot essentially moves along $y$ when driven by the central gate.

\section{Simple model for the variability of the Rabi frequency of electrons}
\label{app:Model}

According to Appendix \ref{app:Rabi}, the Rabi frequency of electrons is $f_R\propto\partial\langle y\rangle/\partial V_{\rm fg}$ in the present setup. In this appendix, we derive a simple model for the variability of $f_R$ in a disordered dot using first-order perturbation theory.

We consider a quantum dot with energies $E_n$ and eigenstates $\ket{\psi_n}$ at zero magnetic field (we discard, therefore, the spin index for simplicity). In the presence of a disorder potential $W(\vec{r})$, the first-order ground state reads:
\begin{equation}
\ket{\tilde{\psi}_0}=\ket{\psi_0}+\sum_{n>0}\frac{\bra{\psi_n}W\ket{\psi_0}}{E_0-E_n}\ket{\psi_n}+{\cal O}(W^2)\,,
\end{equation}
so that:
\begin{equation}
\langle y\rangle=\bra{\psi_0}y\ket{\psi_0}+2\sum_{n>0}\frac{\bra{\psi_0}y\ket{\psi_n}\bra{\psi_n}W\ket{\psi_0}}{E_0-E_n}+{\cal O}(W^2)\,.
\end{equation}
We have assumed real wave functions (as always possible at zero magnetic field in the absence of spin-orbit coupling). Next,
\begin{align}
\langle y\rangle^\prime&=2\bra{\psi_0^\prime}y\ket{\psi_0}+2\sum_{n>0}\frac{1}{E_0-E_n}\Big\{ \nonumber \\
&\bra{\psi_0^\prime}y\ket{\psi_n}\bra{\psi_n}W\ket{\psi_0}+\bra{\psi_0}y\ket{\psi_n^\prime}\bra{\psi_n}W\ket{\psi_0} \nonumber \\
&\bra{\psi_0}y\ket{\psi_n}\bra{\psi_n^\prime}W\ket{\psi_0}+\bra{\psi_0}y\ket{\psi_n}\bra{\psi_n}W\ket{\psi_0^\prime} \nonumber \\
&-\frac{E_0^\prime-E_n^\prime}{E_0-E_n}\bra{\psi_0}y\ket{\psi_n}\bra{\psi_n}W\ket{\psi_0} \Big\}+{\cal O}(W^2)\,,
\end{align}
where $f^\prime$ stands for $\partial f/\partial V_{\rm fg}$. Moreover,
\begin{subequations}
\begin{align}
E_n^\prime&=-e\bra{\psi_n}D\ket{\psi_n} \\
\ket{\psi_n^\prime}&=-e\sum_{m\ne n}\frac{\bra{\psi_m}D\ket{\psi_n}}{E_n-E_m}\ket{\psi_m}\,,
\end{align}
\end{subequations}
where, as before, $D(\vec{r})=\partial V(\vec{r})/\partial V_{\rm fg}$ is the derivative of the electrostatic potential $V(\vec{r})$ in the device with respect to the front gate voltage $V_{\rm fg}$.

This problem can be solved analytically in some paradigmatic cases. We assume that the motions along $x$, $y$, $z$ are separable in the pristine dot, and that the confinement along $y$ is harmonic, with characteristic energy $\hbar\omega_y$. We also assume that the gate creates a homogeneous electric field along $y$, that is $D(\vec{r})=y/L$ with $L$ some characteristic device length. Then, using the relation
\begin{equation}
\bra{i}y\ket{j}=\sqrt{\frac{\hbar}{2m_\parallel^*\omega_y}}\left(\sqrt{j+1}\delta_{i,j+1}+\sqrt{j}\delta_{i,j-1}\right)
\end{equation}
for the eigenstates $\ket{i}$ of the harmonic oscillator, we get:
\begin{equation}
\langle y\rangle^\prime=\frac{e}{m_\parallel^*\omega_y^2L}\left(1+F[W]\right)\,,
\end{equation}
where:
\begin{equation}
F[W]=\frac{1}{\hbar\omega_y}\int d^3\vec{r}\,W(\vec{r})\rho(\vec{r})\,,
\end{equation}
and:
\begin{equation}
\rho(\vec{r})=\psi_0(\vec{r})^2-\psi_1(\vec{r})^2-\sqrt{2}\psi_0(\vec{r})\psi_2(\vec{r})\,.
\end{equation}
Here $\psi_1(\vec{r})$ and $\psi_2(\vec{r})$ are the wave functions with respectively one and two quanta of excitation along $y$. Therefore, if $E[W(\vec{r})]=0$, the average Rabi frequency $\overline{f_R}=f_R^0$ is that of the pristine device, and:
\begin{equation}
\tilde{\sigma}(f_R)=\tilde{\sigma}(\langle y\rangle^\prime)=\sigma(F)\,,
\end{equation}
where:
\begin{equation}
\sigma^2(F)=\frac{1}{(\hbar\omega_y)^2}\int d^3\vec{r}\int d^3\vec{R}\,S(\vec{r})\rho(\vec{R})\rho(\vec{R}+\vec{r})\,,  
\end{equation}
with $S(\vec{r})=E[W(\vec{R})W(\vec{R}+\vec{r})]$ the auto-correlation function of the disorder (assumed independent on $\vec{R}$). We next introduce the Fourier transform of $S(\vec{r})$ (the power spectrum of the disorder):
\begin{equation}
S(\vec{q})=\int d^3\vec{r}\,S(\vec{r})e^{-i\vec{q}\cdot\vec{r}}\,,
\end{equation}
and reach:
\begin{equation}
\tilde{\sigma}^2(f_R)=\frac{1}{(\hbar\omega_y)^2}\int \frac{d^3\vec{q}}{(2\pi)^3}S(\vec{q})\left|\rho(\vec{q})\right|^2\,.
\label{eq:model}
\end{equation}

We now apply this expression to interface roughness disorder in a thin film with thickness $H$. When $H\ll\ell_{\mathcal{E}_z}$, the disorder potential $W(\vec{r})$ is given by Eq. (\ref{eq:WSR}), and the auto-correlation function $S(\vec{r})$ is twice Eq. (\ref{eq:autoSR}) (because there are two independent top and bottom interfaces). Since $W(\vec{r})$ depends only on $\vec{r}_\parallel$, we just need to specify the wave function in that plane, and to integrate Eq. (\ref{eq:model}) over $\vec{q}_\parallel$ [$d^3\vec{q}/(2\pi)^3\to d^2\vec{q}_\parallel/(2\pi)^2]$. Therefore, we also assume parabolic confinement along $x$, with a possibly different characteristic energy $\hbar\omega_x$, so that: 
\begin{align}
\rho(\vec{r}_\parallel)&=\frac{1}{\pi\ell_x^0\ell_y^0}\exp\left(-\frac{x^2}{2(\ell_x^0)^2}\right)\exp\left(-\frac{y^2}{2(\ell_y^0)^2}\right) \nonumber \\
&\times\left(1-\frac{y^2}{(\ell_y^0)^2}\right)\,,
\label{eq:rho}
\end{align}
where $\ell_x^0=\sqrt{\hbar/(2m_\parallel^*\omega_x)}$ and $\ell_y^0=\sqrt{\hbar/(2m_\parallel^*\omega_y)}$ are the characteristic sizes of the pristine dot. Substituting the Fourier transforms of Eqs. (\ref{eq:autoSR}) and (\ref{eq:rho}) into Eq. (\ref{eq:model}), we finally reach:
\begin{align}
\tilde{\sigma}(f_R)&=\sqrt{6}\pi^2\frac{m_\parallel^*}{m_\perp^*}\frac{\Delta L_c\ell_y^0}{H^3}\left(\frac{\ell_y^0}{\ell_x^0}\right)^{1/2}\nonumber \\
&\times\left(\frac{4(\ell_x^0)^2}{4(\ell_x^0)^2+L_c^2}\right)^{1/4}\left(\frac{4(\ell_y^0)^2}{4(\ell_y^0)^2+L_c^2}\right)^{5/4}
\label{eq:harmonic}
\end{align}
This expression highlights several key points discussed in this work:
\begin{itemize}
    \item The variability is proportional to the rms fluctuations $\Delta$ (see Figs. \ref{fig:SRHoles} and  \ref{fig:SRElectrons}), and scales as $\partial E_\perp/\partial H \propto 1/(m_\perp^*H^3)$ with the thickness of the film and the confinement mass of the carriers [Eq. (\ref{eq:WSR})].
    \item The variability has a non-monotonous behavior with $L_c$, and shows a peak at:
    \begin{subequations}
    \begin{align}
    L_c&=\frac{1}{\sqrt{2}}\Big(\sqrt{(\ell_y^0)^4+26(\ell_x^0\ell_y^0)^2+9(\ell_x^0)^4} \nonumber \\
    &+(\ell_y^0)^2-3(\ell_x^0)^2\Big)^{1/2}\,, \\
    &\simeq2\sqrt{\frac{2}{3}}\ell_y^0\text{ when }\ell_x^0\gg\ell_y^0\,, \\
    &\simeq\ell_y^0\text{ when }\ell_y^0\gg\ell_x^0\,. 
    \end{align}
    \end{subequations}
    As discussed in section \ref{sec:Surfaceroughness}, fluctuations with wave lengths $\lambda\ll\ell_y^0$ are averaged out, while the interface becomes flat again on the scale of the dot when $\lambda\gg\ell_y^0$. The fluctuations with wave lengths comparable to the size of the dot are actually the most detrimental to the variability.
    \item At given dot sizes $\ell_x^0$ and $\ell_y^0$, the variability is proportional to the in-plane mass $m_\parallel^*$: the heavier the particles, the stronger they localize in the disorder. We emphasize nonetheless that $\ell_x^0$ and $\ell_y^0$ also depend on $m_\parallel^*$ for a given confinement potential.
    \item When increasing the dot size $\ell_x^0$ or $\ell_y^0$, the dot becomes more polarizable and responsive to the disorder. Yet fluctuations along the driving axis $y$ are expected to have much more impact on the motion of the dot than fluctuations along the transverse axis $x$. Actually, $\tilde{\sigma}(f_R)$ even decreases monotonously with increasing $\ell_x^0$, because the fluctuations get better averaged out on the scale of the dot. However, $\tilde{\sigma}(f_R)$ increases monotonously with increasing $\ell_y^0$; it scales as $(\ell_y^0)^4$ when $\ell_y^0\ll L_c$, and as $(\ell_y^0)^{3/2}$ when $\ell_y^0\gg L_c$. The relative variability $\tilde{\sigma}(f_R)$ decays slower than the absolute variability $\sigma(f_R)$ when $\ell_y^0\to0$ because $\sigma(f_R)$ also tends to 0 as $(\ell_y^0)^4$.
\end{itemize}

We point out that the expression of $\tilde{\sigma}(f_R)$ as a function of $\ell_x^0$, $\ell_y^0$ and $L_c$ is distinctive of the confinement potential (e.g., parabolic $vs$ triangular), of the operators coupling the spin and electric fields (e.g., $y$ for electrons and $k_y^2$ for holes), and on the power spectrum $S(\vec{q})$ of the disorder. In the devices investigated in this work, the dots are confined in an anharmonic potential along $y$, and show weaker scalings with $\ell_y^0$ (see Fig. \ref{fig:SRHoles}) than expected from Eq. (\ref{eq:harmonic}), especially for holes.
    
Finally, we briefly discuss the dependence of the variability of $f_R$ on the density $n_i$ of $P_b$ defects at the Si/SiO$_2$ interface. As a prototypical example, we consider an ensemble of $N$ defects at random positions $\{\vec{R}_{i\parallel}\}$ on a planar interface ${\cal S}$ with area $S$ located at $z=0$. We assume that the vertical confinement at this interface is strong enough that we can introduce an effective 2D potential $\nu(\vec{r}_\parallel,\vec{R}_\parallel)$ for a point charge at position $\vec{R}_\parallel$ (the defect potential averaged over the ground-state envelope function along $z$). We also assume that the interface is homogeneous, so that $\nu(\vec{r}_\parallel, \vec{R}_\parallel)\equiv\nu(\vec{r}_\parallel-\vec{R}_\parallel)$ only depends on $\vec{r}_\parallel-\vec{R}_\parallel$.\footnote{We emphasize that $\nu(\vec{r}_\parallel)$ decays typically faster than $1/|\vec{r}_\parallel|$ due to the presence of metal gates in the device.} Then, the total disorder potential reads:
\begin{equation}
W(\vec{r}_\parallel)=\sum_{i=1}^N \nu(\vec{r}_\parallel-\vec{R}_{i\parallel})\,.
\end{equation}
If the $\vec{R}_{i\parallel}$ are uncorrelated, we reach when $S\to\infty$:\cite{Niquet14}
\begin{subequations}
\begin{align}
E[W(\vec{r}_\parallel))]&=n_i\tilde{\nu} \\
S(\vec{r}_\parallel)&=n_i S_c(\vec{r}_\parallel)+n_i^2\tilde{\nu}^2\,,
\end{align}
\end{subequations}
where:
\begin{equation}
\tilde{\nu}=\int_{\cal S} d^2\vec{R}_\parallel\,\nu(\vec{R}_\parallel)\,,
\end{equation}
and:
\begin{equation}
S_c(\vec{r}_\parallel)=\int_{\cal S} d^2\vec{R}_\parallel\,\nu( \vec{R}_\parallel)\nu(\vec{R}_\parallel+\vec{r}_\parallel)
\end{equation}
is the spatial auto-correlation function of the potential of a single charge. $E[W(\vec{r}_\parallel)]$ is the potential created by a homogeneous density of traps at the interface that simply shifts the dot energies, while only $S_c(\vec{r})$ gives rise to variability. Therefore, $\tilde{\sigma}(f_R)$ is expected to scale as $\sqrt{n_i}$ according to Eq. (\ref{eq:model}). Indeed, two defects would scatter Rabi frequencies twice as much as a single defect only if they were systematically at the same positions.

\begin{figure}
\centering
\includegraphics[width=.95\columnwidth]{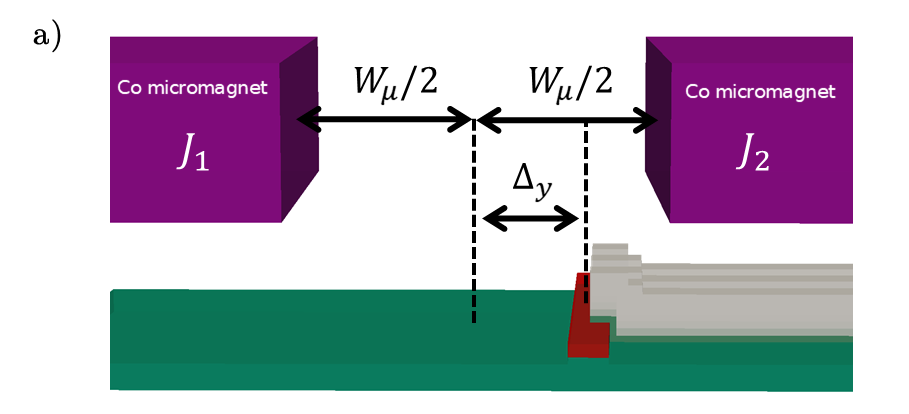}
\includegraphics[width=.95\columnwidth]{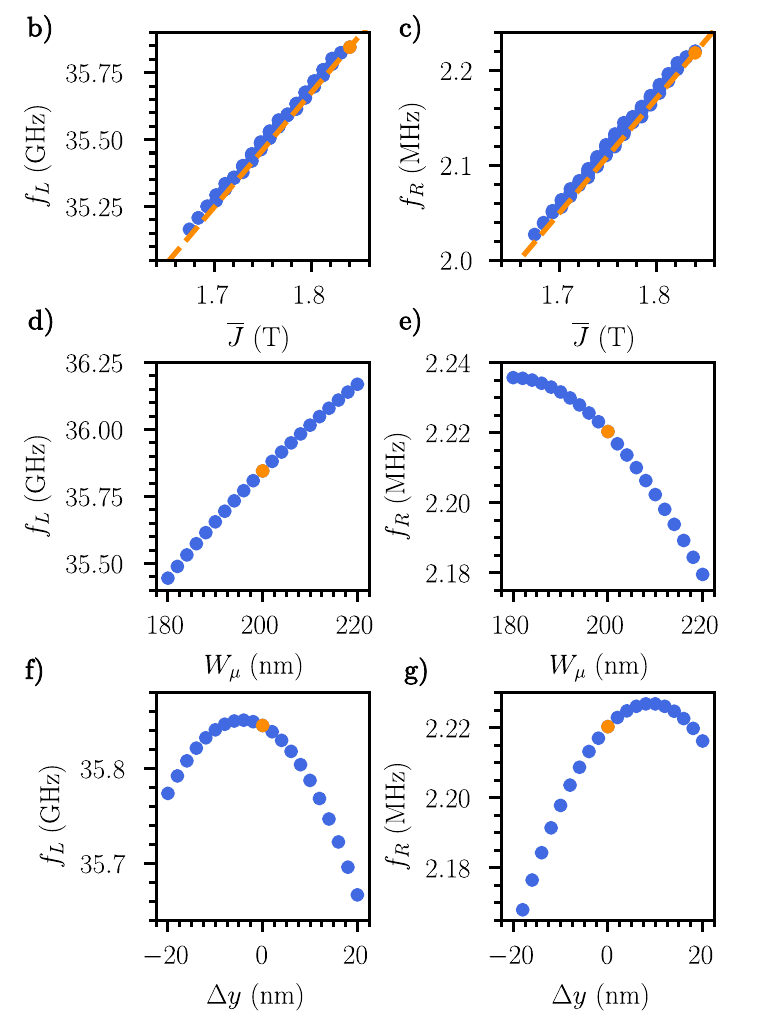}
\caption{\YMN{(a) Sketch of the device with the definition of the width $W_\mu$ and misalignment $\Delta_y$ of the trench between the micro-magnets. (b) Larmor and (c) Rabi frequency of the pristine electron qubit as a function of the average magnetic polarisation $\overline{J}=(J_1+J_2)/2$ of the two micro-magnets. The plots collect data for different $\Delta J=J_1-J_2$ ranging from $-0.17$ to $0.17$ T. The orange lines are simple linear models $f_R\propto\overline{J}$ and $f_L\propto\overline{J}+J_0$, where $J_0$ accounts for the static magnetic field. (d) Larmor and (e) Rabi frequency as a function of the width $W_\mu$ of the trench between the two magnets. (f) Larmor and (g) Rabi frequency as a function of the misaligment $\Delta y$ between the channel and the trench. In all panels, $V_{\rm fg}=50$ mV and the orange point is the nominal device ($\overline{J}=1.84$ T, $W_\mu=200$ nm, and $\Delta y=0$).}}
\label{fig:micromagnetdisorder}
\end{figure}

The data of Figs. \ref{fig:CTholes1} and \ref{fig:CTelectrons} scale faster than $\sqrt{n_i}$ but slower than $n_i$. We attribute this discrepancy to the breakdown of the above assumptions. In particular, the average Rabi frequency $\overline{f_R}$ decreases with increasing $n_i$ in the non-parabolic confinement potential of the 1D channel, which strengthen the scaling of $\tilde{\sigma}(f_R)$. Also, the variability is dominated by the defects that are close to or within the dot, and whose effects go beyond first-order perturbation theory. At small densities, the likelihood to have a defect within the dot directly scales as $n_i$.

\section{Effects of micro-magnet imperfections}
\label{app:Varmicro}

\YMN{In this Appendix, we briefly discuss the effects of disorder in the micro-magnets on the variability of electron spin qubits. The relevant parameters of the micro-magnets are displayed in Fig. \ref{fig:micromagnetdisorder}a.}

\YMN{The qubits may be sensitive to local inhomogeneities of the magnets (roughness, variations of the magnetic polarization $J$), and to ``global'' (but more systematic) deficiencies such as misalignment (misplacement and misorientation).\cite{Yoneda15,Simion20} The roughness of the magnets tends to be softened in the far field and is likely not a strong concern, unless particularly large or long-ranged. The variations of the magnetic polarization $J$ due to material inhomogeneity or incomplete saturation can be readily addressed when they take place over length scales much longer than the distance to the qubits (that is, in the hundreds of nm range). The magnetic polarization $J$ can then be considered as locally homogeneous, but device dependent. The Larmor and Rabi frequencies of the pristine qubit are thus plotted in Figs. \ref{fig:micromagnetdisorder}b,c as a function of the average magnetic polarisation $\overline{J}=(J_1+J_2)/2$ of the two magnets (they are almost independent on $\Delta J=J_1-J_2$ in this range). The Rabi frequency being directly proportional to the gradient of the micro-magnets field, any relative variation of $\overline{J}$ results in a similar relative variation of $f_R$ (dotted lined $f_R\propto\overline{J}$ in Fig. \ref{fig:micromagnetdisorder}c). The Larmor frequency $f_L$ shows a weaker, yet significant linear dependence on $\overline{J}$ as the micro-magnets field is only $\simeq 20\%$ of the the total magnetic field (dotted lined in Fig. \ref{fig:micromagnetdisorder}b). As a matter of fact, a $1.3\%$ variation of $\overline{J}$ results in a 100 MHz drift of the Larmor frequency, which is sizable with respect to the distributions shown in Figs. \ref{fig:SRElectrons} and \ref{fig:CTelectrons}. The homogeneity of the magnets can, therefore, be critical for the control of the Larmor frequencies.}

\begin{figure}
\centering
\includegraphics[width=.9\columnwidth]{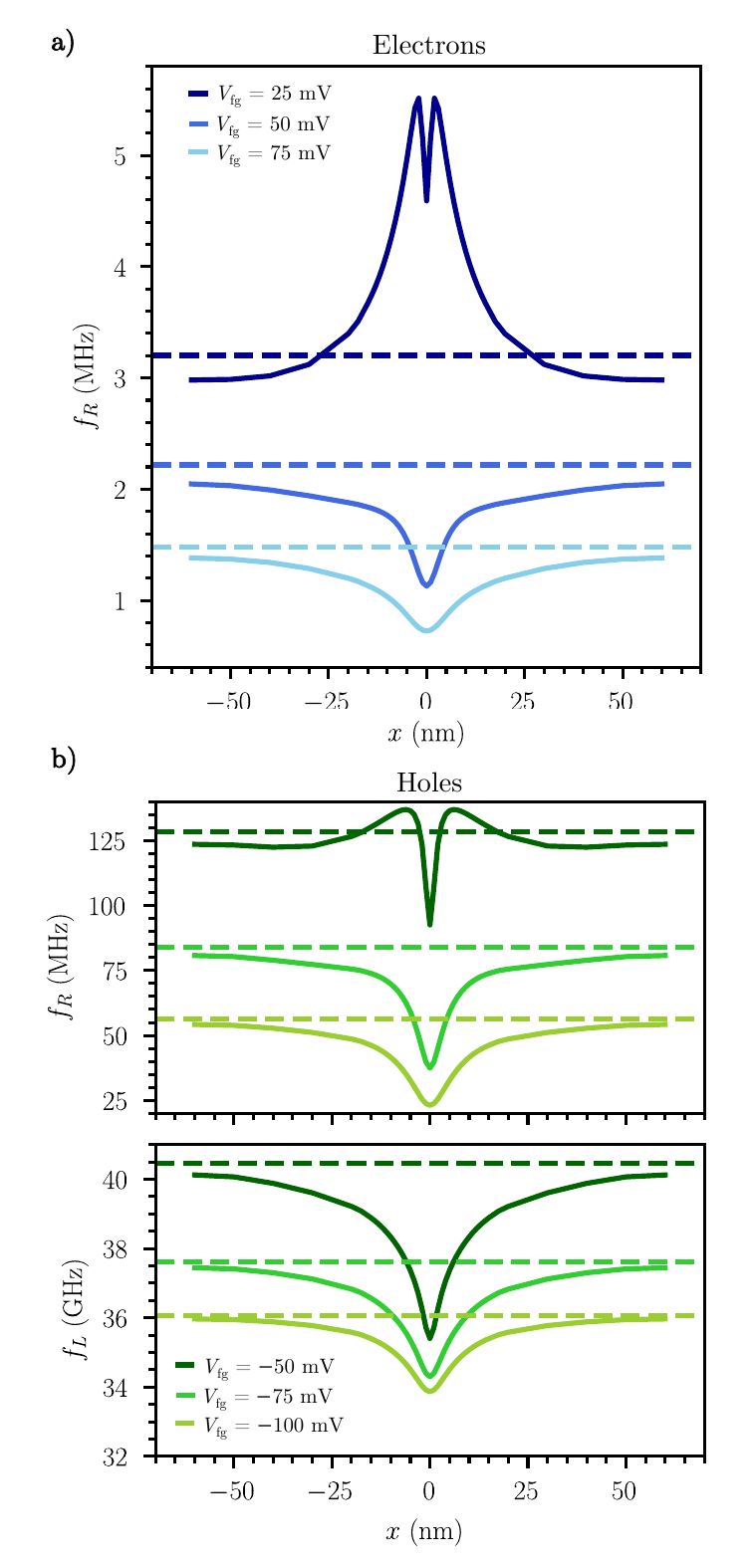}
\caption{(a) Rabi frequency of electron qubits as a function of the position $x$ of a single negative charge along the channel, for different gate voltages $V_{\rm fg}$. The charge is located on the top Si/SiO$_2$ interface, at $y=10$ nm. The horizontal lines are the Rabi frequencies of the pristine qubits. (b) Rabi and Larmor frequency of electron qubits as a function of the position $x$ of a single positive charge along the channel, for different gate voltages $V_{\rm fg}$. The charge is located on the top Si/SiO$_2$ interface, at $y=0$ nm. The horizontal lines are the frequencies of the pristine qubits.} 
\label{fig:singlecharge}
\end{figure}

\YMN{We have also plotted in Figs. \ref{fig:micromagnetdisorder}d,e the Larmor and Rabi frequencies as a function of the width $W_\mu$ of the trench between the magnets (nominally $W_\mu=200$ nm). $f_L$ increases and $f_R$ decreases when widening the trench because $B_y$ increases (the qubit looks better aligned with the magnets) but $\partial B_z/\partial y$ decreases. Making a bevel trench is actually a solution to detune the qubits on purpose in order to address them individually at different Larmor frequencies ($\partial f_L/\partial W_\mu\approx 18$ MHz/nm).\cite{Yoneda15} Finally, the Larmor and Rabi frequencies are plotted in Figs. \ref{fig:micromagnetdisorder}f,g as a function of the misaligment $\Delta y$ between the channel and the micro-magnets trench. The variations are small and mostly second-order in the $|\Delta y|<20$ nm range because symmetric positions on both sides of the $(xz)$ mirror plane of the magnets are roughly (but not strictly) equivalent (the major component $B_y$ is the same but the minor component $B_z$ changes sign). If the micro-magnets are misoriented by $2^\circ$ with respect to the channel axis, neighboring qubits (that are 60 nm apart in the design of Fig. \ref{fig:device}) are shifted by $\Delta y=\pm 2$ nm.}

\section{Additional figures}
\label{app:Additions}

In this Appendix, we provide additional data on the effects of charge traps on the Larmor and Rabi frequencies of electron and hole qubits, and discuss some results on hole qubits driven in the iso-Zeeman EDSR mode.

\subsection{Effects of a single charge on electron and hole qubits}

The Larmor and Rabi frequency of electron and hole qubits is plotted in Fig. \ref{fig:singlecharge} as a function of the position $x$ of a single charge along the channel. This charge is positive for hole qubits, negative for electron qubits, and is located on the top facet. The deviations from the pristine qubit are sizable when the charge is within $\approx 25$ nm from the qubit. Note the ``overshoot'' of the Rabi frequency at small bias when the charge goes through the gate and tends to split the dot in two strongly coupled pieces.

\begin{figure}
\centering
\includegraphics[width=.95\columnwidth]{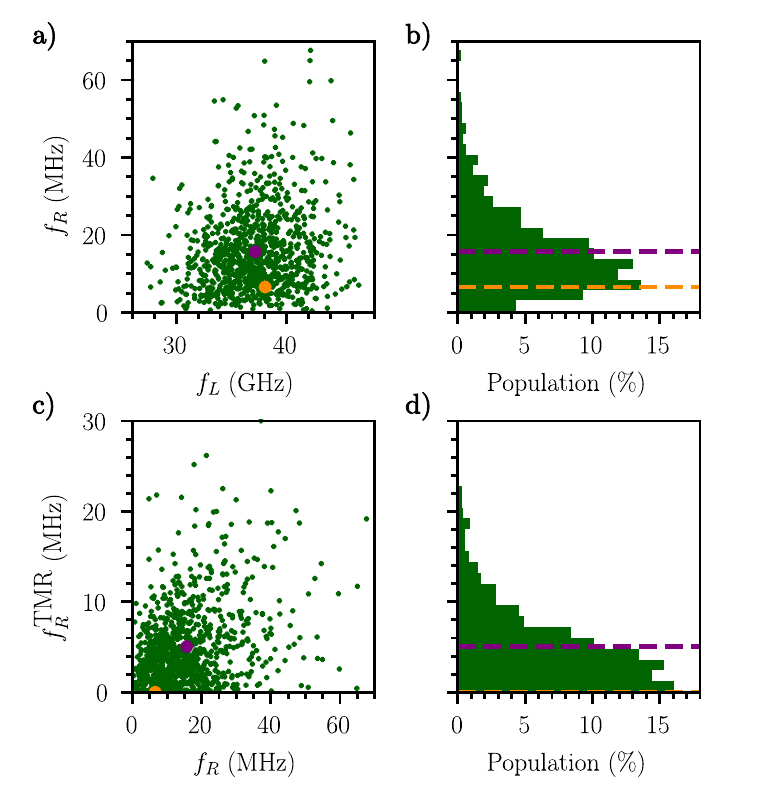}
\caption{(a) Distribution of the Larmor and Rabi frequencies, and (b) histogram of the Rabi frequencies of rough hole devices driven in the iso-Zeeman EDSR mode ($\Delta=0.4$ nm, $L_c=10$ nm and $V_{\rm fg}=-50$ mV). The pure $g$-TMR contribution $f_R^{\rm TMR}$ arising from the sole modulation of the principal $g$-factors is plotted as a function of the total Rabi frequency in (c), and its histogram in (d). Each green point of panels (a) and (c) is a particular realization of the interface roughness disorder. The orange points and lines are the pristine device, while the purple dots and lines are the average device.} 
\label{fig:IZ_SR}
\end{figure}

\begin{figure}
\centering
\includegraphics[width=.95\columnwidth]{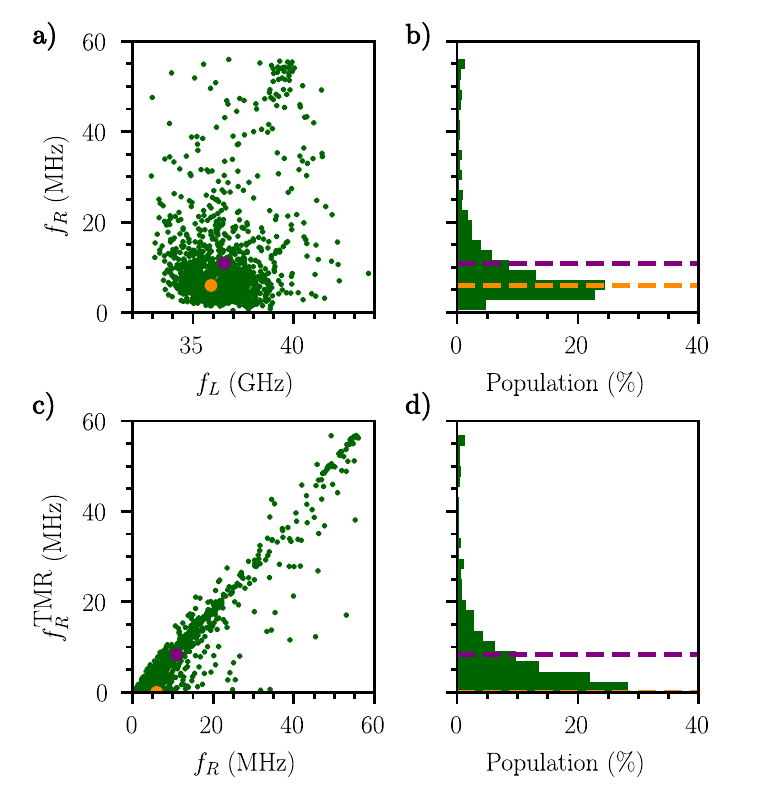}
\caption{(a) Distribution of the Larmor and Rabi frequencies, and (b) histogram of the Rabi frequencies of hole devices with charge traps driven in the iso-Zeeman EDSR mode ($n_i=5\times 10^{10}$ cm$^{-2}$, $V_{\rm fg}=-50$ mV). The pure $g$-TMR contribution $f_R^{\rm TMR}$ arising from the sole modulation of the principal $g$-factors is plotted as a function of the total Rabi frequency in (c), and its histogram in (d). Each green point of panels (a) and (c) is a particular realization of the interface roughness disorder. The orange points and lines are the pristine device, while the purple dots and lines are the average device.} 
\label{fig:IZ_CT}
\end{figure}

\begin{figure}
\centering
\includegraphics[width=.95\columnwidth]{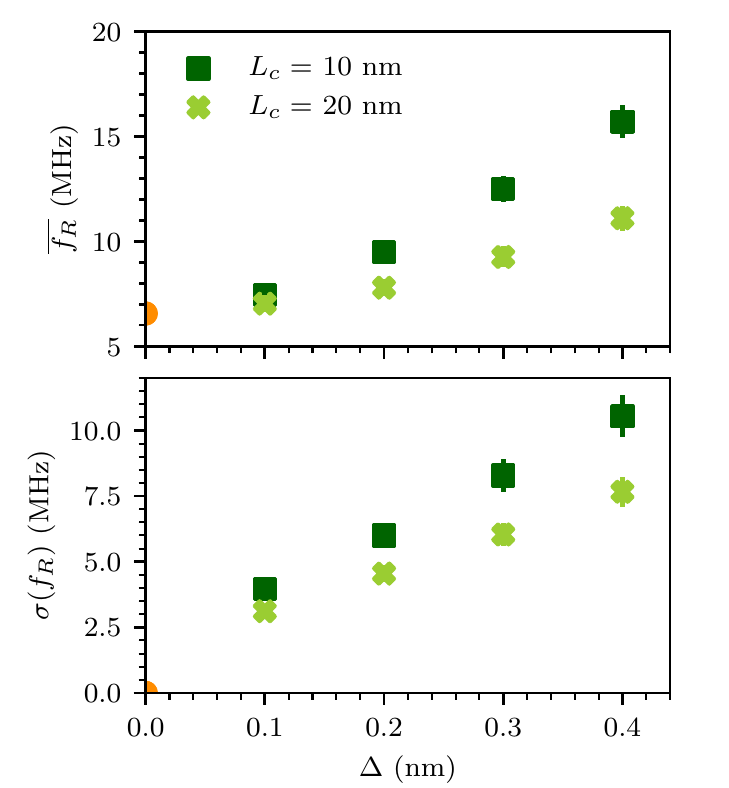}
\caption{Average Rabi frequency $\overline{f_R}$, and absolute standard deviation $\sigma(f_R)$ as a function of the rms interface roughness $\Delta$ for two correlation lengths $L_c=10$ and $20$ nm. The orange dot is the pristine device.}
\label{fig:IZ_DELTA}
\end{figure}

\subsection{Hole qubits driven in the iso-Zeeman EDSR mode}

We now consider a hole qubit driven by a RF field $V_{\rm l}(t)=V_{\rm RF}\sin(2\pi f_L t)/2$ on the left gate and $V_{\rm r}(t)=-V_{\rm RF}\sin(2\pi f_L t)/2$ on the right gate. This field shakes the dot as a whole in the quasi-harmonic confinement potential along the channel axis; the principal $g$-factors are therefore little dependent on the position $\langle x\rangle$ of the hole (no $g$-TMR in the pristine device). The dot however experiences Rashba-type spin-orbit interactions that rotate the spin.\cite{Kloeffel11,Kloeffel18,Michal21} The magnetic field $\vec{B}_{\rm ext}$ is oriented along $\vec{y}-\vec{z}$, which is closer to the optimum in this configuration.

At variance with the $g$-TMR mode, the Rabi frequency of the pristine device increases with increasingly negative $V_{\rm fg}$ because the Rashba spin-orbit coupling gets enhanced by the lateral and vertical electric fields. It reaches $f_R\approx 7$ MHz at $V_{\rm fg}=-100$ mV (where it is almost saturated).\footnote{The static bias on the left and right gates is set to $V_{\rm l}=V_{\rm r}=-75$ meV in order not to impede the motion of the dot by a stiff confinement along $x$.} It is much smaller than the $g$-TMR Rabi frequency because $g$-tensor modulation is more efficient than Rashba spin-orbit interactions over a wide range of bias voltages in such heavy-hole devices.\cite{Michal21} Moreover, the left and right gates are pretty far from the qubit in the present layout (and are partly screened by the front gate), which can be however compensated by a larger drive amplitude.

The distribution of Larmor and Rabi frequencies for interface roughness ($\Delta=0.4$ nm, $L_c=10$ nm) and charge traps ($n_i=5\times 10^{10}$ cm$^{-2}$) are plotted in Figs. \ref{fig:IZ_SR} and \ref{fig:IZ_CT}. The RSD $\tilde{\sigma}(f_R)=67.13\%$ for interface roughness and $\tilde{\sigma}(f_R)=100.28\%$ for charge traps are much larger than in the $g$-TMR mode (respectively $\tilde{\sigma}(f_R)=24.44\%$ and $\tilde{\sigma}(f_R)=63.07\%$). Strikingly, the average Rabi frequency systematically increases in the presence of disorder. This results from the emergence of a strong $g$-TMR contribution on top of the original iso-Zeeman mechanism in the disordered qubits (the gyromagnetic factors now depending significantly on the position of the dot along $x$). To support this conclusion, we have plotted the distribution of this $g$-TMR contribution, calculated from the sole dependence of the principal $g$-factors on the driving RF field, in Figs. \ref{fig:IZ_SR} and \ref{fig:IZ_CT}. It is indeed zero in the pristine qubit, but can reach up to a few tens of MHz in the disordered devices. $g$-TMR even becomes the prevalent mechanism in a significant fraction of the qubits in the presence of charge traps (as the wave function can be substantially distorted by the RF drive). The weak confinement along $x$, necessary for efficient driving (the iso-Zeeman Rabi frequency scales as $\ell_x^4$),\cite{Michal21} as well as the mixing with a strong, disorder-induced $g$-TMR contribution explain the huge variability of the Rabi frequencies. The average Rabi frequency $\overline{f_R}$, and the absolute standard deviation $\sigma(f_R)$ are also plotted as a function of the rms interface roughness $\Delta$ in Fig. \ref{fig:IZ_DELTA}. As expected, $\sigma(f_R)$ increases almost linearly with strengthening $\Delta$, while $\overline{f_R}$ increases quadratically (since this increase is second order in the disorder when the latter averages to zero, as suggested by Appendix \ref{app:Model}). These trends were confirmed with atomistic tight-binding\cite{Bourdet18} calculations on selected devices.

%\bibliography{Variability}


\end{document}